\documentclass[a4paper, 12pt]{article}
\usepackage[bottom=2cm, top=1.5cm, left=2cm, right=2cm]{geometry}
\usepackage{amsmath,amssymb,amsfonts,amsthm,hyperref,bbm,latexsym}

\usepackage{bbold}

\usepackage{graphicx,multirow, multicol}
\usepackage{cases}
\usepackage{cancel}
\usepackage{url}
\usepackage{indentfirst}
\usepackage[compat=1.1.0]{tikz-feynman}

\usepackage{orcidlink}

\usepackage[width=\textwidth]{caption}
\usepackage[english]{babel}
\usepackage{csquotes}

\usepackage[
backend=biber,
defernumbers=false,
citestyle=numeric-comp,
bibstyle=chem-angew,
sorting=none,
sortcites=true,
doi=true,
eprint=true, 
url=false, 
giveninits=true
]{biblatex} 
\defbibentryset{thooftveltman}{thooft1971a, thooft1971, thooft1972a, thooft1972}
\defbibentryset{almasy2009err}{almasy2009a, almasy2010}
\defbibentryset{dienerkniehl}{barroso2000, diener2001}
\defbibentryset{FeynCalc}{mertig1991, shtabovenko2016, shtabovenko2017, shtabovenko2020, shtabovenko2025}
\defbibentryset{LoopTools}{oldenborgh1990,hahn1999}
\defbibentryset{GrimusNeufeld}{grimus1989,drauksas2019, dudenas2019,dudenas2019a,dudenas2022,dudenas2023,drauksas2024}
\numberwithin{equation}{section}
\allowdisplaybreaks

\begin{document}

\title{Renormalization of mixing angles\\ and\\ computation of the hadronic $W$ decay widths}

\author{
  Simonas Draukšas~\orcidlink{0000-0003-4796-2760}\thanks{E-mail:
    \href{mailto:simonas.drauksas@ff.vu.lt}{simonas.drauksas@ff.vu.lt}}
  \\
  \small Vilnius University, Faculty of Physics, Institute of Theoretical Physics and Astronomy, \\
  \small Saulėtekio av. 9, Vilnius, Lithuania, LT-10222
}

\date{}

\maketitle

\begin{abstract}
We provide a practical prescription for a variant of the On-Shell scheme which does not require mixing matrix counterterms at all, i.e. $\delta V=0$. The scheme is based on the fact that one can always choose a basis in which there are no mixing matrices (angles) and, therefore, the corresponding counterterms are superfluous. Importantly, the prescription is model- and process-independent and is formulated entirely in terms of self-energies. As an example, we compute the 1-loop hadronic $W$--boson decay widths in the Standard Model with different renormalization schemes of the quark mixing matrix found in the literature and the one found in this paper. For full consistency, the principles of this scheme are employed both for the quark mixing matrix and for the Weinberg angle.
\end{abstract}

\section{Introduction}

The ideas and methods of renormalization of the Standard Model (SM) seem to be well established even in the case of particle mixing~\cite{thooftveltman} and are by now material for textbooks and reviews, e.g.~\cite{denner2020}. Nonetheless, the road to the renormalization of the quark mixing matrix in the case of three generations proved to be somewhat bumpy. The Cabibbo--Kobayashi--Maskawa (CKM) matrix~\cite{cabibbo1963,kobayashi1973} was renormalized in the `traditional' On-Shell (OS)~\cite{aoki1982} scheme already more than 30 years ago in~\cite{denner1990}, yet the renormalization proved to introduce unwanted gauge dependence (in $R_\xi$ gauges) in physical $W$ decay amplitudes~\cite{gambino1999}. This gauge dependence sparked quite a few proposals of getting rid of it~\cite{gambino1999,yamada2001,dienerkniehl,zhou2003,denner2004,liao2004,kniehl2006,kniehl2009a} and more recently in~\cite{drauksas2023}. Even more so, by now not only correct UV cancellations and gauge-independence are required from the renormalization of mixing, but also flavor symmetry, physical process independence and numerical stability are desirable~\cite{freitas2002,denner2018}.

In slightly more detail, each successor of the scheme presented in~\cite{denner1990}, except for~\cite{drauksas2023}, aims at the removal of gauge dependence in the mixing matrix counterterms, while keeping all the other traditional On-Shell (field, mass, and coupling) counterterms~\cite{aoki1982}. The authors of~\cite{gambino1999} propose to define the CKM counterterm by using an additional set of field renormalization counterterms defined from self-energies at 0 momentum transfer. An arbitrary separation of the gauge-dependent and gauge-independent parts was proposed in~\cite{liao2004}, one inspired by the pinch technique in~\cite{yamada2001}, and one inspired by the structure of the external-leg corrections in~\cite{kniehl2006} with a version formulated in terms of the self-energies in a flavor-democratic way~\cite{kniehl2009a}. The CKM matrix counterterm in~\cite{dienerkniehl} is derived by comparing to a theory without particle mixing. The scheme of~\cite{zhou2003} takes a two-step approach by first setting the relevant form factor in the $W\to ud$ decay amplitude to 0 and then requiring for the unitarity of the bare CKM matrix. The scheme in~\cite{zhou2003} is streamlined and reproduced in~\cite{denner2004}, where symmetry of the SM is discussed and a physical renormalization condition is proposed. All of these schemes define a gauge-independent CKM counterterm, but not all of these counterterms can be called On-Shell or, at least, not in the traditional sense. For example, the initial scheme of~\cite{denner1990} defines the mixing matrix counterterm fully in terms of traditional On-Shell field counterterms and can be called On-Shell even if that leads to gauge-dependence. The scheme in~\cite{gambino1999} is not an OS scheme due to self-energies evaluated at 0 momentum. The schemes of~\cite{yamada2001,dienerkniehl,liao2004,kniehl2006,kniehl2009a} actually imply an \textit{additional} set of field renormalization counterterms different to that of traditional OS counterterms in order to define a gauge-independent CKM counterterm, therefore, these schemes are at least not traditional OS schemes. The schemes of~\cite{zhou2003} and~\cite{denner2004} do not have an additional set of field counterterms, although, these schemes impose an arbitrary renormalization condition on a relevant form factor of the hadronic $W$ decay amplitude (process-dependence). To be fair, there does not really exist a natural renormalization condition for mixing matrices (perhaps due to degeneracy of field renormalization and mixing matrix counterterms~\cite{altenkamp2017,drauksas2023a}), for example, even in~\cite{denner1990} the authors are forced to make their definition by the UV finiteness of the amplitude, therefore, the schemes in~\cite{zhou2003} and~\cite{denner2004} are as On-Shell as~\cite{denner1990}. Shortly put, the discussed schemes either do not pass some of the mixing renormalization requirements, or are not consistently On-Shell.

In this paper, the main focus is given to the scheme in~\cite{drauksas2023}, which we have developed previously and which we find to be a consistent On-Shell scheme that also passes all the mixing renormalization requirements. The scheme is based on the same traditional OS renormalization conditions, but they are fulfilled by a set of counterterms with a trivial mixing matrix counterterm and non-diagonal mass counterterms. In spirit this scheme is very similar to the ones of~\cite{kniehl2006,kniehl2009a}, but in~\cite{drauksas2023} we have argued against the diagonalization of mass counterterms, which would give rise to mixing matrix counterterms. In our view, such a diagonalization violates basis invariance and is responsible for quite a few problems, which have been outlined in~\cite{drauksas2023a} and encountered already in~\cite{freitas2002}. In addition, the diagonalization of the mass counterterm implies two sets of field renormalization counterterms, which we have avoided for consistency. The principles of basis invariance can be applied to the Weinberg angle counterterm, in turn, this counterterm is also trivial in our approach. 

A part of the analytical work for our scheme in~\cite{drauksas2023} has already been done, but a really practical prescription was lacking. In this paper we develop such a prescription, which is a universal and simple formula for the calculation of the anti-hermitian part of the field renormalization entirely in terms of self-energies such that the mixing matrix counterterm is trivial. In addition, we provide the first numerical results for this scheme. We do so in the spirit of~\cite{almasy2009err}, where the numerical impact for the $W^+\to u\bar{d}$ decay widths in the Standard Model for various CKM renormalization schemes is discussed. We aim to extend the Table I found in that paper with the results of our scheme. In doing so, we show that our scheme is plausible not only analytically, but also numerically. 

The structure of the paper is the following. In Section~\ref{sec:RenoBig} we provide the detailed construction of renormalization in the gauge and quark sectors with special attention to the renormalization of mixing angles (matrices). Our main result is the practical prescription for the anti-hermitian part of the field renormalization in Section~\ref{sec:method}. In Section~\ref{sec:Wreno} we arrive at the renormalized $W$ boson decay width into quarks and in Section~\ref{sec:numres} we provide the numerical results for the partial and total hadronic $W$ decay widths with different renormalization schemes for the quark mixing matrix: namely, the schemes of Refs.~\cite{denner1990,gambino1999,diener2001,kniehl2006}, $\overline{\mathrm{MS}}$ scheme, a scheme with no mixing on external legs~\cite{almasy2009err}, and our scheme in~\cite{drauksas2023} on which we focus on in this work. In Section~\ref{sec:conclusions} we give our conclusions and there are two appendices.

\section{Renormalization of the hadronic \texorpdfstring{$W$}{W} decay at 1-loop}
\label{sec:RenoBig}
\subsection{Renormalization of mixing angles}
\label{sec:anglereno}

Mixing angles are derived parameters coming from the preference to work in a basis in which the mass matrix is diagonal. Mixing angles are then treated as independent parameters which are to be renormalized, despite the fact that they are an artifact of basis change. One can notice this fact by a rather simple procedure: take a multiplet of real massless scalar fields $\hat{\phi}_i$, a multiplet of real scalar fields $\hat{\chi}^\prime_j$ with a non-diagonal mass matrix, and introduce a diagonal interaction between these with another real scalar field, say $\hat{\Phi}$. Then the relevant part of the Lagrangian is as follows
\begin{equation}
  \mathcal{L}= 
              -\frac{1}{2}\sum_{j,j^\prime} \hat{\chi}^\prime_j (\hat{M}^2_{\chi})_{jj^\prime}^{}\hat{\chi}^\prime_{j^\prime}
              -\lambda \sum_{i}\hat{\Phi} \hat{\phi}_i\hat{\chi}^\prime_i\,.
\end{equation}
Here the hats indicate bare quantities. We can change the basis of the fields $\hat{\chi}^\prime_j=\hat{U}_{jj^\prime}\hat{\chi}_{j^\prime}$ such that the mass matrix $\hat{U}^T \hat{M}^2_\chi \hat{U}=\hat{m}_\chi^2$ is diagonal after the orthogonal transformation. While the transformation matrix can be easily included in the mass matrix, it remains in the interaction term

\begin{equation}
\mathcal{L}= 
            -\frac{1}{2}\sum_{j} \hat{\chi}_j (\hat{m}^2_\chi)_{jj}^{}\hat{\chi}_{j}
            -\lambda \sum_{i,j}\hat{\Phi} \hat{\phi}_i \hat{U}_{ij}\hat{\chi}_j\,.
\end{equation}
This remaining matrix of basis transformation is then called a mixing matrix and is usually renormalized as $\hat{U}_{ij}=U_{ij}+\delta U_{ij}$ (the renormalization of other parameters and fields is irrelevant at the moment, so we keep them bare for simplicity)
\begin{equation}
\mathcal{L}= 
            -\frac{1}{2}\sum_{j} \hat{\chi}_j (\hat{m}^2_{\chi})_{jj}^{}\hat{\chi}_{j}
            -\lambda \sum_{i,j}\hat{\Phi} \hat{\phi}_i \left( U_{ij}+\delta U_{ij} \right) \hat{\chi}_j\,.
\end{equation}
It is important to note that basis transformations must remain available in the renormalized version, hence, we are free to transform the field $\chi_j$ by the inverse of $U$, i.e. $\hat{\chi}_j=U^{T}_{jj^\prime}\hat{\chi}^{\prime\prime}_{j^\prime}$, then we have
\begin{equation}
\begin{split}
    \mathcal{L}=& 
            -\frac{1}{2}\sum_{i^\prime,j^\prime} \hat{\chi}^{\prime\prime}_{i^\prime} (\hat{M}^2_{\chi^{\prime\prime}})_{i^\prime j^\prime}^{}\hat{\chi}^{\prime\prime}_{j^\prime}
            -\lambda \sum_{i,j^\prime}\hat{\Phi} \hat{\phi}_i \left(\delta_{ij^\prime}+\left(\delta U U^{T} \right)_{ij^\prime} \right) \hat{\chi}^{\prime\prime}_{j^\prime}\,,
\end{split}
\end{equation}
where we have defined $\hat{M}^2_{\chi^{\prime\prime}}=U \hat{m}_\chi U^T$, which is non-diagonal. Importantly, the counterterm $\delta \hat{U}=\delta U U^{T}$ remains and is in general non-vanishing even though there is no mixing matrix anywhere else in the Lagrangian so that the counterterm does not have an associated parameter. It follows that for consistency with basis rotations the mixing angle (matrix) counterterms should not be introduced, i.e. the bare mixing matrix should be identified with the renormalized one
\begin{equation}
U=\hat{U} \quad \Rightarrow \quad \delta U=0\,.
\end{equation}
Shortly put, basis rotations should not be renormalized as they are not related to physical quantities. 

The non-physicality can be seen in yet another way. Rotate the fields with the bare mass matrix and then renormalize them, then to first order we have
\begin{equation}
  \label{eq:degeneracy}
  \hat{\chi}^\prime=\hat{U}Z\chi=U\left(I+\delta Z + U^T\delta U\right)\chi\,.
\end{equation}
Here it can be seen that the (anti-symmetric part of the) field renormalization counterterms are degenerate with the mixing matrix counterterms~\cite{altenkamp2017,drauksas2023a}. The non-physicality of the field counterterms then immediately transfers to the mixing matrix counterterms. Simultaneously, it means that all the effects of the basis change can be absorbed into field renormalization. This degeneracy has also been noticed in Ref.~\cite{denner2004}, with the conclusion that an On-Shell prescription for the mixing matrix counterterm is impossible. In addition, this degeneracy implies that the counterterm $\delta U$ can be safely absorbed into the field renormalization, i.e. set to 0.

Yet another way to see that a trivial counterterm, $\delta U=0$, is a valid choice, consider that all the eigenvalues of mixing matrices are equal to 1, which should not have counterterms. Of course, this is obscured in a different basis, but this is still one of the distinguishing features of mixing matrices. For example, mass matrices have eigenvalues, which are genuine parameters of the theory.

On the other hand, mixing itself is a physical phenomenon and should be renormalized, but renormalization of mixing matrices singles out a basis, which is not consistent. Instead, the proposal is to renormalize mass matrices
\begin{equation}
  \hat{M}^2_{ij}=M^2_{ij}+\delta M^2_{ij}
\end{equation}
and diagonalize just the renormalized part
\begin{equation}
\label{eq:massdiag}
\begin{split}
    \left( U^T \hat{M}^2 U \right)_{ij}
            =\left( U^T M^2 U \right)_{ij}+\left( U^T\delta M^2U \right)_{ij}
            =m^2_{ii}\delta_{ij}+\delta m^2_{ij}\,.
\end{split}
\end{equation}
Importantly, even with a diagonal renormalized mass the mass counterterm $\delta m^2$ is in general non-diagonal and removes the need for mixing matrix counterterms. Note that there is and should not be any difference, whether one first does a basis rotation and renormalizes, or renormalizes and then diagonalizes since basis rotations should always be possible without changing the results. This is ensured if there are no counterterms associated with basis rotations. However, note that the mixing angles can still be used to parametrize the theory, but they are always renormalized and do not have counterterms in our approach.

Having briefly laid out the main idea --- introduce off-diagonal mass counterterms to not have mixing angle counterterms --- in the following subsections we apply it to the relevant sectors of the Standard Model, namely, the electroweak (EW) sector and the quark sector. We present the application in detail as in some certain steps it differs from the standard approaches.

\subsection{Renormalization of the Standard Model electroweak sector and the Weinberg angle}
\label{sec:gaugereno}

\subsubsection{Setup of counterterms in the electroweak sector}

We begin with a rather simple case of the Standard Model electroweak (EW) sector. The covariant derivative is 
\begin{equation}
  D_\mu = \partial_\mu - i \hat{g} T^a \hat{W}^a_\mu - i \hat{g}^\prime Y \hat{B}_\mu\,,
\end{equation}
where $\hat{g}$ and $\hat{g}^\prime$ are the $SU(2)_L$ and $U(1)_Y$ gauge couplings, respectively, $T^a=\tau^a/2$ are the $SU(2)_L$ generators with $a$ being the adjoint index and $\tau^a$ the Pauli matrices in isospin space, $Y$ is the $U(1)_Y$ generator (the hypercharge), $\hat{W}^a_\mu$ and $\hat{B}_\mu$ are the respective gauge fields, and the hats continue to indicate bare quantities. Mass terms of the gauge bosons are generated via the Spontaneous Symmetry Breaking (SSB) mechanism by the vacuum expectation value $\hat{v}$ of the Higgs doublet $\hat{H}$ ($SU(2)_L$ doublet with hypercharge $Y=1/2$)
\begin{equation}
  \hat{H}=\begin{pmatrix}
      \hat{G}^+ \\
      \frac{1}{\sqrt{2}}(\hat{v}+\hat{h}+i\hat{G}^0)
    \end{pmatrix}\,,
\end{equation}
where $\hat{h}$ is the physical Higgs boson and $\hat{G}^+$ and $\hat{G}^0$ are the would-be Goldstone bosons. The kinetic term of the Higgs doublet at the minimum produces the well known mass matrix for the gauge bosons
\begin{equation}
\hat{M}^2_{VV^\prime}=\begin{pmatrix}
  \hat{M}^2_{WW} & 0_{2\times2} \\
  0_{2\times2} & \hat{M}^2_{W B}
\end{pmatrix}\,,  
\end{equation}
with the $2\times2$ diagonal block for the $\hat{W}^{1,2}$ bosons
\begin{equation}
\label{eq:bareWmass}
  \hat{M}^2_{WW}= \frac{\hat{g}^2 \hat{v}^2}{4} I = \hat{m}^2_W I\,,
\end{equation}
with $I$ being the identity matrix, and another $2\times2$ block for the $\hat{W}^3$ and $\hat{B}$ bosons
\begin{equation}
  \label{eq:WBblock}
  \hat{M}^2_{W B}=\frac{\hat{v}^2}{4}\begin{pmatrix}
    \hat{g}^2 & -\hat{g} \hat{g}^\prime \\
    -\hat{g} \hat{g}^\prime & \hat{g}^{\prime 2}
  \end{pmatrix}=\begin{pmatrix}
    \hat{m}^2_W & -\hat{m}_B \hat{m}_W \\
    -\hat{m}_B \hat{m}_W & \hat{m}^2_B
  \end{pmatrix}\,.
\end{equation}

The standard step would be to diagonalize $\hat{M}^2_{W B}$ by an orthogonal transformation, however, we first renormalize the mass matrices
\begin{equation}
  \hat{M}^2_{WW} = M^2_{WW}+\delta M^2_{WW}\,, \quad \hat{M}^2_{W B} = M^2_{W B}+\delta M^2_{W B}  
\end{equation}
or, equivalently,
\begin{equation}
  \hat{m}^2_W= m^2_W+\delta m^2_W\,, \quad
  \hat{m}^2_B= m^2_B+\delta m^2_B\,.
\end{equation}
To first order the counterterm matrix $\delta M^2_{W B}$ is

\begin{equation}
  \delta M^2_{W B}=\begin{pmatrix}
    \delta m^2_W & -\frac{m_W}{2m_B} \delta m^2_B - \frac{m_B}{2m_W} \delta m^2_W \\
    -\frac{m_W}{2m_B} \delta m^2_B - \frac{m_B}{2m_W} \delta m^2_W & \delta m^2_B
  \end{pmatrix}\,.
\end{equation}

Diagonalization of the now renormalized mass matrix $M^2_{W B}$ by $R$ in analogy with Eq.~\eqref{eq:massdiag} relates the $W^3$ and $B$ fields to the mass eigenstates, which are the photon $A$ and the $Z$ boson (this relation holds both for bare and renormalized fields)
\begin{equation}
\label{eq:W3Brotation}
  \begin{pmatrix}
    W^{3}_\mu\\B_\mu
  \end{pmatrix}
  =R\begin{pmatrix}
    Z_\mu\\A_\mu
  \end{pmatrix}
  =\begin{pmatrix}c & s\\
  -s & c
  \end{pmatrix}
  \begin{pmatrix}
    Z_\mu\\A_\mu
  \end{pmatrix}\,,
\end{equation}
where the Weinberg angle $\theta$ is defined as usual
\begin{equation}
  c\equiv \cos\theta = \frac{m_W}{m_Z}\,, \qquad s\equiv \sin\theta =\sqrt{1-c^2} 
\end{equation}
and the $Z$ boson mass is
\begin{equation}
\label{eq:Zmassren}
  m_Z^2 = m^2_B+m^2_W\,.
\end{equation}

The same transformation by $R$ leads to 
\begin{equation}
  R^T \delta M^2_{WB} R = \begin{pmatrix}
    \delta m^2_Z & \delta m^2_{ZA} \\
    \delta m^2_{AZ} & 0
  \end{pmatrix}\,,
\end{equation}
with 
\begin{equation}
\label{eq:Zmassct}
  \delta m^2_Z= \delta m^2_B+\delta m^2_W
\end{equation}
and
\begin{equation}
  \label{eq:AZmassct}
  \delta m^2_{AZ}=\delta m^2_{ZA}=\frac{1}{2sc}\left(\delta m_{W}^{2}-c^{2} \delta m_{Z}^{2}\right)\,.
\end{equation}

It is worth noting that the photon is massless and the corresponding mass counterterm is zero, as expected due to gauge invariance. The Weinberg angle is defined entirely in terms of renormalized masses and does not require a counterterm. It is rather simple to see that the counterterm $\delta m^2_{AZ}$ replaces the usual counterterm for the Weinberg angle as the two are related by
\begin{equation*}
  \frac{\delta m^2_{AZ}}{m^2_Z} =-\frac{\delta \sin \theta_W}{\sin\theta_W}\,,
\end{equation*}
as can be seen by comparing with the definitions found, for example, in~\cite{denner2020}. Having discussed the inconsistency of mixing angle counterterms, we do not use $\delta \sin \theta_W$ and work only with $\delta m^2_{AZ}$.

Returning to Eq.~\eqref{eq:WBblock} and renormalizing the gauge couplings as
\begin{equation}
  \hat{g} = g + \delta g\,, \qquad \hat{g}^\prime = g^\prime + \delta g^\prime\,,
\end{equation}
we may note that the same rotation by $R$ also diagonalizes the matrix of renormalized gauge couplings, so that
\begin{equation}
  \label{eq:ggprime}
  c = \frac{g}{\sqrt{g^2+g^{\prime 2}}}\,, \qquad gs=g^\prime c\,,
\end{equation}
and one may find the relation between the gauge coupling and $\delta m^2_{AZ}$ counterterms
\begin{equation}
\label{eq:ggprimerelation}
  \frac{\delta g}{g}-\frac{\delta g^\prime}{g^\prime}=\frac{1}{cs}\frac{\delta m^2_{AZ}}{m^2_Z}\,.
\end{equation}

The counterterms $\delta g/g$ and $\delta g^\prime/g^\prime$ can be fixed by considering the covariant derivative in the broken phase and considering gauge invariance (charge universality). The renormalized and relevant $Z-A$ part of the covariant derivative can be written as follows

\begin{equation}
  \label{eq:AZderivative0}
  \begin{split}
    D^{Z-A}_\mu=&
     -i\begin{pmatrix}
    \hat{g}~T^3 & \hat{g}^\prime~Y
  \end{pmatrix}
  \cdot
  \begin{pmatrix}
    c & s \\
    -s & c
  \end{pmatrix}
  \cdot
    \begin{pmatrix}
    \hat{Z}_\mu \\
    \hat{A}_\mu
  \end{pmatrix}
    \\  
  =&-i\begin{pmatrix}
    (g+\delta g) T^3 & (g^\prime + \delta g^\prime) Y
  \end{pmatrix}
  \cdot
  \begin{pmatrix}
    c & s \\
    -s & c
  \end{pmatrix}
  \cdot
  \begin{pmatrix}
    1+\delta Z_{ZZ} & \delta Z_{ZA} \\
    \delta Z_{AZ} & 1+\delta Z_{AA}
  \end{pmatrix}
  \cdot
  \begin{pmatrix}
    Z_\mu \\
    A_\mu
  \end{pmatrix}\,,
  \end{split}
\end{equation}

where we have put the gauge couplings together with the corresponding generators in a row vector, used Eq.~\eqref{eq:W3Brotation}, and with $\delta Z_{VV^\prime}$ being the field renormalization counterterms of the neutral gauge bosons. Note that $\delta Z_{VV^\prime}$ adheres to basis rotations and can be related to the field renormalization counterterms of the fields $W^3$ and $B$ through a transformation by $R$. Multiplying everything out, separating the photon and $Z$ boson parts and defining the electric charge generator 
\begin{equation}
  Q=T^3+Y\,,
\end{equation}
to first order we have

\begin{equation}
\label{eq:AZderivative}
  \begin{split}
    Z_\mu&:-i \frac{g}{c}\left( T^3-Q s^2 +Qs^2 \left( \frac{c}{s}\delta Z_{AZ}-\frac{\delta g^\prime}{g^\prime}-\delta Z_{ZZ} \right) +T^3\left( \delta Z_{ZZ}+s^2\frac{\delta g^\prime}{g^\prime}+c^2\frac{\delta g}{g} \right)\right)\,, \\
    A_\mu&: -igs\left(Q+Q\left(\delta Z_{AA}-\frac{s}{c}\delta Z_{ZA}+\frac{\delta g^\prime}{g^\prime}\right) +T^3\left( \frac{\delta g}{g}-\frac{\delta g^\prime}{g^\prime}+\frac{1}{sc}\delta Z_{ZA}\right)\right)\,.
  \end{split}
\end{equation}

Here we have used Eq.~\eqref{eq:ggprime} as needed and exchanged $Y$ for $Q-T^3$ in order to see the vector and axial--vector structure, since both left- and right-handed particles have the same electric charge $Q$ (by construction of the model), while fields have different $T^3$ depending on handedness. The only massless gauge boson after SSB is the photon, which couples irrespective of handedness, therefore, the photon terms proportional to $Q$ in Eq.~\eqref{eq:AZderivative} correspond to the $U(1)_{EM}$ renormalized covariant derivative of the broken phase.

Gauge invariance requires that the covariant derivative of a field transforms like the field itself under gauge transformations. Hence, if renormalization is to preserve gauge invariance, the covariant derivative of a field must be renormalized like the field itself, that is
\begin{equation}
  D_\mu \hat{\psi} = Z_\psi D_\mu \psi\,,
\end{equation} 
for some arbitrary field $\psi$ with field renormalization constant $Z_\psi$. In turn, it must be true that

\begin{equation}
  \left( \partial_\mu - i gs Q A_\mu \right)Z_\psi \psi = \left(\partial_\mu -i  gs Q\left(1+\delta Z_{AA}-\frac{s}{c}\delta Z_{ZA}+\frac{\delta g^\prime}{g^\prime}\right)A_\mu  \right)Z_\psi \psi\,.
\end{equation}
 
The above equality requires that the counterterms in the parentheses on the r.h.s. cancel out. Therefore, we get for the renormalization of the $U(1)_Y$ coupling
\begin{equation}
\label{eq:gprimect}
  \frac{\delta g^\prime}{g^\prime} = -\delta Z_{AA}+\frac{s}{c}\delta Z_{ZA}
\end{equation}
and by using Eq.~\eqref{eq:ggprimerelation} we arrive at a counterterm for the  $SU(2)_L$ coupling
\begin{equation}
\label{eq:gct}
  \frac{\delta g}{g} = -\delta Z_{AA}+\frac{s}{c}\delta Z_{ZA}+\frac{1}{c s}\frac{\delta m^2_{AZ}}{m^2_Z}\,.
\end{equation}

Following A.~Sirlin~\cite{sirlin1980} we may also define the bare electric charge by rewriting the covariant derivative in Eq.~\eqref{eq:AZderivative0} in terms of $Q$ and $Q_Z=T^3-Qs^2$
\begin{equation}
  \begin{split}
    D^{Z-A}_\mu=&-iA_\mu\left[
      Q\left(\hat{g}s^3+\hat{g}^\prime c^3\right)
      +Q_Z\left(\hat{g}s-\hat{g}^\prime c\right) \right]
    -iZ_\mu \left[
      Qsc\left(\hat{g}s-\hat{g}^\prime c\right)
    +Q_Z\left(\hat{g}c+\hat{g}^\prime s\right)\right]\,.
  \end{split}
\end{equation}
The bare electric charge is then the combination of bare couplings where the photon couples to the electromagnetic current, hence
\begin{equation}
  \hat{e}\equiv\hat{g}s^3+\hat{g}^\prime c^3\,.
\end{equation}
This gives the usual relation
\begin{equation}
  \label{eq:eggprime}
  e=gs=g^\prime c
\end{equation}
and by using Eqs.~\eqref{eq:gprimect} and~\eqref{eq:gct} one finds the following electric charge counterterm 
\begin{equation}
  \label{eq:ect}
  \begin{split}
      \frac{\delta e}{e}=&s^2\frac{\delta g}{g}+c^2\frac{\delta g^\prime}{g^\prime}=
      -\delta Z_{AA}
      +\frac{s}{c}\delta Z_{ZA}
      +\frac{s}{c}\frac{\delta m^2_{AZ}}{m^2_Z}\,.
  \end{split}
\end{equation}

Interestingly, in our approach, the counterterm of $g^\prime$ looks identical to the counterterm of electric charge found in the literature, e.g.~\cite{denner1993,pokorski2000,bohm2001,denner2020}, while $\delta e$ is unusual. However, this is only in appearance, since the definition of $\delta Z_{ZA}$ is different from the usual On-Shell approach due to the presence of off-diagonal mass counterterm $\delta m^2_{AZ}$ as we see in the section below. In turn, once all the field and mass counterterms are written explicitly in terms of self-energies, the coupling counterterms $\delta g$, $\delta g^\prime$, and $\delta e$ are as in traditional On-Shell scheme~\cite{aoki1982}, see Eqs.~\eqref{eq:gprimectSE},~\eqref{eq:gctSE}, and~\eqref{eq:ectSE} below. This means that, for example, the renormalized electric charge is as measured in the Thomson limit. 

For completeness, we give the mass of the $Z$ boson, which can be retrieved from the rotated bare matrices in Eq.~\eqref{eq:WBblock},
\begin{equation}
\label{eq:bareZmass}
  \hat{m}_Z^2 = m_Z^2 + \delta m_Z^2= \frac{\hat{v}^2}{4}\left(\hat{g} c + \hat{g}^\prime s\right)^2\,.
\end{equation}
Note that from this equation or Eq.~\eqref{eq:bareZmass} we could define a counterterm for the vacuum expectation value. However, this is not necessary as we always `absorb' the vev into the masses and include tadpole diagrams in the self-energy calculations. This amounts to using the so-called Fleisher--Jegerlehner tadpole scheme~\cite{fleischer1981,denner2016}, which ensures gauge-independent mass counterterms.

Finally, the $W^{1,2}$ bosons are rotated into their ($Q$) charge eigenstates as usual, so that
\begin{equation}
  W^\pm_\mu = \frac{1}{\sqrt{2}}\left( W^1_\mu \mp i W^2_\mu \right)\,.
\end{equation}
The $W^{\pm}$ bosons are then renormalized as
\begin{equation}
  \hat{W}_\mu^{\pm} = Z_W W_\mu^{\pm}\,.
\end{equation}
Note that for field renormalization constants we do not use the square root convention as already evident from the renormalization of the photon and $Z$ fields in Eq.~\eqref{eq:AZderivative0}.

\subsubsection{Two point functions and renormalization conditions in the electroweak sector}

Having set up all the counterterms, we may write down the renormalized two-point functions for the gauge bosons of the Standard Model. We may split every self-energy $\Pi^{\mu\nu}_{VV^\prime} (p^2)$ into transverse $\Pi_{VV^\prime}(p^2)$ and longitudinal $\Pi^{L}_{VV^\prime}(p^2)$ parts as
\begin{align}
        i\Pi^{\mu\nu}_{V^\prime V} (p^2)=&
        \vcenter{\hbox{\begin{tikzpicture}
        \begin{feynman}
        \vertex [blob] (m1) at (0,0){1PI};
        \vertex (a) at (-1.4,0.0) ;
        \vertex (b) at (1.4,0);
        \diagram*{
        (a)--[photon, edge label=$V$](m1)--[photon, edge label=$V^\prime$](b);
        };
        \end{feynman}
        \end{tikzpicture}}}
        \hspace{1em}
        +
        \hspace{1em}
        \vcenter{\hbox{\begin{tikzpicture}
        \begin{feynman}
        \vertex [blob] (m1) at (0,0.75){1PI};
        \vertex (a) at (-1.4,0.0) ;
        \vertex (b) at (1.4,0);
        \vertex (c) at (0,0);
        \diagram*{
        (a)--[photon, edge label=$V$](c)--[photon, edge label=$V^\prime$](b);
        (c)--(m1);
        };
        \end{feynman}
        \end{tikzpicture}}}
        \nonumber \\
        & \nonumber \\
        =& \left(g^{\mu\nu}-\frac{p^\mu p^\nu}{p^2}\right)i\Pi_{V^\prime V}(p^2)+\frac{p^\mu p^\nu}{p^2} i\Pi^{L}_{V^\prime V}(p^2)\,, 
\end{align}
where we assume that the loop functions are real (we implicitly use the $\widetilde{Re}$ operator~\cite{denner1993}, which drops the so-called absorptive parts of loop functions.).

The renormalized transverse parts of the two-point functions $\Pi^R_{V^\prime V}$ for the SM gauge bosons are
\begin{align}
  \Pi^R_{WW}(p^2)=&\Pi_{WW}(p^2)-2\left(p^2-m_W^2\right)\delta Z_W + \delta m_W^2\,, \\
  \Pi^R_{ZZ}(p^2)=&\Pi_{ZZ}(p^2)-2\left(p^2-m_Z^2\right)\delta Z_{ZZ} + \delta m_Z^2\,, \\
  \Pi^R_{AA}(p^2)=&\Pi_{AA}(p^2)-2p^2\delta Z_{AA}\,, \\
  \Pi^R_{AZ}(p^2)=&\Pi_{AZ}(p^2)-\delta Z^T_{AZ} \left( p^2-m^2_Z \right)
  -p^2\delta Z_{AZ} + \delta m_{AZ}^2\,.
  \nonumber
\end{align}

Proceeding with the traditional On-Shell renormalization conditions~\cite{aoki1982}, we require that the renormalized masses are the pole masses, i.e. $\Pi^R_{VV}(m_V^2)=0$, which fixes the mass counterterms of the $Z$ and $W$ bosons. This leads to the following counterterms
\begin{equation}
  \delta m_W^2 =-\Pi_{WW}(m_W^2)\,, \quad \delta m_Z^2 =-\Pi_{ZZ}(m_Z^2)\,.
\end{equation}
As we have included tadpole diagrams in the self-energies, these mass counterterms are gauge-independent~\cite{gambino2000}. Consequently, via Eq.~\eqref{eq:AZmassct},  these mass counterterms also produce the off-diagonal mass counterterm
\begin{equation}
\label{eq:AZmassctfinal}
  \delta m_{AZ}^2 = -\frac{1}{2sc}\left(\Pi_{WW}(m_W^2)-c^2 \Pi_{ZZ}(m_Z^2) \right)\,.
\end{equation}
 The tadpoles cancel in this counterterm, hence, it is gauge-independent irrespective of the tadpole treatment.

The unit residue requirement leads to $\partial_{p^2} \Pi^R_{VV}\vert_{p^2=m_V^2}=0$ and results in the following field renormalization counterterms
\begin{equation}
\label{eq:diagonalfieldct}  
    \begin{split}
  \delta Z_W =& \frac{1}{2}\left. \frac{\partial \Pi_{WW}(p^2)}{\partial p^2} \right\vert_{p^2=m_W^2}\,, \\
  \delta Z_{ZZ} =& \frac{1}{2} \left. \frac{\partial \Pi_{ZZ}(p^2)}{\partial p^2} \right\vert_{p^2=m_Z^2}\,, \\
  \delta Z_{AA} =& \frac{1}{2}\left. \frac{\partial \Pi_{AA}(p^2)}{\partial p^2} \right\vert_{p^2=0}\,.
  \end{split} 
\end{equation}
Lastly, the propagators should be diagonal on-shell, hence, the $Z-A$ self-energy at $p^2=\{m_Z^2,0\}$ should vanish, which leads to
\begin{equation}
\label{eq:offdiagonalfieldct}
\begin{split}
    \delta Z_{AZ} =& \frac{1}{m_Z^2}\left( \Pi_{AZ}(m_Z^2) + \delta m^2_{AZ} \right)\,, \\ 
  \delta Z^T_{AZ}=&\delta Z_{ZA} = -\frac{1}{m_Z^2} \left( \Pi_{AZ}(0) + \delta m^2_{AZ} \right)\,.
\end{split}
\end{equation}
It is also worth noting the symmetric and antisymmetric parts of the $Z-A$ field renormalization constants
\begin{align}
  \delta Z^S_{AZ} &= \frac{1}{2}\left(\delta Z_{AZ}+\delta Z^T_{AZ}\right)
  = \frac{1}{2m_Z^2}\left( \Pi_{AZ}(m_Z^2)-\Pi_{AZ}(0) \right)\,,\\
  \delta Z^A_{AZ} &= \frac{1}{2}\left(\delta Z_{AZ}-\delta Z^T_{AZ}\right)
  = \frac{1}{2m_Z^2}\left( \Pi_{AZ}(m_Z^2)+\Pi_{AZ}(0)+2\delta m^2_{AZ} \right)\,.
\end{align}
Here the sign change under transposition is ensured by noting that the denominator $1/m_Z^2$ is actually a difference of squared masses $1/(m_Z^2-m_A^2)$ with $m_A=0$, which changes sign under transposition. There is a general feature: the off-diagonal mass counterterms are related with the anti-symmetric (anti-hermitian) part of the field renormalization constants, while the symmetric (hermitian) part is independent of mass counterterms~\cite{altenkamp2017,drauksas2023a}. It is also interesting to note that in our approach the anti-symmetric part of the field renormalization is UV finite as can be checked explicitly. One can find that the UV divergences contained in the bare self-energies cancel against the ones of $\delta m^2_{AZ}$ in the expression for $\delta Z^A_{AZ}$, which can be employed in the renormalization of the effective leptonic Weinberg angle in the $\overline{\mathrm{MS}}$ scheme~\cite{gambino1994} due to the correspondence between $\delta m^2_{AZ}$ and $\delta \sin \theta_W$.

Having all the field and mass counterterms defined in terms of the self-energies we may employ Eqs.~\eqref{eq:gprimect},~\eqref{eq:gct},~\eqref{eq:AZmassctfinal},~\eqref{eq:diagonalfieldct}, and~\eqref{eq:offdiagonalfieldct} to express the gauge coupling counterterms in terms of self-energies
\begin{align}
  \frac{\delta g^\prime}{g^\prime}=& -\frac{1}{2}\left.\frac{\partial \Pi_{AA}(p^2)}{\partial p^2}\right\vert_{p^2=0}
  -\frac{s}{c}\frac{\Pi_{AZ}(0)}{m^2_Z}
  \label{eq:gprimectSE}
  +\frac{1}{2c^2 m^2_Z}\left(\Pi_{WW}(m_W^2)-c^2 \Pi_{ZZ}(m_Z^2)\right)\,,
  \\
  \frac{\delta g}{g}=& -\frac{1}{2}\left.\frac{\partial \Pi_{AA}(p^2)}{\partial p^2}\right\vert_{p^2=0}
  -\frac{s}{c}\frac{\Pi_{AZ}(0)}{m^2_Z}
  \label{eq:gctSE}
  -\frac{1}{2s^2 m_Z^2}\left(\Pi_{WW}(m_W^2)-c^2 \Pi_{ZZ}(m_Z^2)\right)\,.
\end{align}
Having these gauge coupling counterterms it is not hard to find the electric charge counterterm from Eq.~\eqref{eq:ect}
\begin{equation}
  \label{eq:ectSE}
\begin{split}
  \frac{\delta e}{e}=&-\frac{1}{2}\left.\frac{\partial \Pi_{AA}(p^2)}{\partial p^2}\right\vert_{p^2=0}
  -\frac{s}{c}\frac{\Pi_{AZ}(0)}{m^2_Z}\,,
\end{split}
\end{equation}
which corresponds to the usual On-Shell counterterm~\cite{denner1993,pokorski2000,bohm2001,denner2020} up to minus signs and factors of 2 due to our conventions of self-energies, choice of the rotation $R$ and field renormalization convention.

One can see that the presence of the counterterm $\delta m^2_{AZ}$, i.e. the triviality of the Weinberg angle counterterm, amounts to a relabeling of counterterms in the gauge-sector. This is because the gauge-sector is quite restricted by gauge-invariance and two-point functions are fully sufficient to renormalize the sector. As we see below, the situation is more difficult in the fermion sector, where the Yukawa couplings are not required by gauge-invariance and there is more freedom. 

\subsection{Renormalization of the Standard model quark sector and the CKM matrix}
\label{sec:quarkreno}

\subsubsection{Setup of counterterms in the quark sector}

To renormalize the quarks we only need the kinetic and mass terms of the Lagrangian
\begin{align}
  \mathcal{L}_\text{kin.+mass}^\text{quark}=&
          -i \sum_{u}\overline{\hat{\psi}^\prime_{u}}\gamma^{\mu}D_\mu\hat{\psi}^\prime_{u}
          -\sum_{u, u^\prime}\overline{\hat{\psi}_{u}^\prime}\hat{M}_{uu^\prime}\hat{\psi}^\prime_{u^\prime} 
          \nonumber \\ \label{eq:quarkkinlagflav}
          &-i \sum_{d}\overline{\hat{\psi}^\prime_{d}}\gamma^{\mu}D_\mu\hat{\psi}^\prime_{d}
          -\sum_{d, d^\prime}\overline{\hat{\psi}_{d}^\prime}\hat{M}_{dd^\prime}\hat{\psi}^\prime_{d^\prime}\,,
\end{align}
where $\hat{\psi}_{u,d}$ are the up- and down-type quark fields in flavor basis such that the indices $u$ and $d$ can correspond to any of the flavors, $\hat{M}_{uu^\prime}$ and $\hat{M}_{dd^\prime}$ are the respective quark mass matrices, and the covariant derivative is understood to be written after SSB, so that it includes only photon and gluon terms. We may rotate the quark fields into a different basis via 
\begin{align}
  \hat{\psi}^\prime_u&=\sum_{uu^\prime}\mathcal{U}_{uu^\prime}\hat{\psi}_{u^\prime}=\left( (U_L)_{uu^\prime} P_L+ (U_R)_{uu^\prime} P_R \right)\hat{\psi}_{u^\prime}\,, 
  \nonumber \\
  \hat{\psi}^\prime_d&=\sum_{dd^\prime}\mathcal{D}_{dd^\prime}\hat{\psi}_{d^\prime}=\left( (D_L)_{dd^\prime} P_L+ (D_R)_{dd^\prime} P_R \right)\hat{\psi}_{d^\prime}\,.  
  \label{eq:fermionrotation}
\end{align}
Here $P_{L,R}=\frac{1}{2}\left( 1\mp\gamma^5\right)$ and the rotation matrices $\mathcal{U}$ and $\mathcal{D}$ are identified with the renormalized ones, i.e. their counterterms are trivial. Then Eq.~\eqref{eq:quarkkinlagflav} becomes
\begin{align}
    \label{eq:quarkkinlagmass0}
      \mathcal{L}_\text{kin.+mass}^\text{quark}=&
          -i\sum_{u} \overline{\hat{\psi}_{u}}\gamma^{\mu}D_\mu\hat{\psi}_{u}
          -\sum_{u, u^\prime}\overline{\hat{\psi}_{u}}\hat{m}_{uu^\prime}\hat{\psi}_{u^\prime}
          -i \sum_{d}\overline{\hat{\psi}_{d}}\gamma^{\mu}D_\mu\hat{\psi}_{d}
          -\sum_{d, d^\prime}\overline{\hat{\psi}_{d}}\hat{m}_{dd^\prime}\hat{\psi}_{d^\prime}\,,
\end{align}
where we have defined
\begin{align}
    \sum_{u^{\prime\prime},u^{\prime\prime\prime}}\gamma^0 (\mathcal{U}^\dagger)_{uu^{\prime\prime}}\gamma^0 \hat{M}_{u^{\prime\prime}u^{\prime\prime\prime}} \mathcal{U}_{u^{\prime\prime\prime}u^\prime} &= \hat{m}_{uu^\prime}\,, 
    \nonumber\\
    \sum_{d^{\prime\prime},d^{\prime\prime\prime}}\gamma^0 (\mathcal{D}^\dagger)_{d^{\prime\prime}}\gamma^0 \hat{M}_{d^{\prime\prime}d^{\prime\prime\prime}} \mathcal{D}_{d^{\prime\prime\prime}d^\prime} &= \hat{m}_{dd^\prime}\,.
    \label{eq:massrotation}
\end{align}
The bare mass matrices $\hat{m}$ are renormalized in analogy with Eq.~\eqref{eq:massdiag}
\begin{equation}
\begin{split}
  \hat{m}_{uu^\prime} &= m_{u} \delta_{uu^\prime}+\delta m_{uu^\prime}\,,\\
  \hat{m}_{dd^\prime} &= m_{d} \delta_{dd^\prime}+\delta m_{dd^\prime}\,.
\end{split}
\end{equation}
Since the renormalized masses $m$ are diagonal, we call this the mass eigenstate basis and the indices $u$ and $d$ are to be understood as any of the following sets $\{u\,,c\,,t\}$ and $\{d\,,s\,,b\}$, respectively. Hermiticity of the Lagrangian imposes a pseudo-hermiticity relation for the mass matrices
\begin{equation}
  \gamma^0  \hat{M} \gamma^0 = \gamma^0 \left( \hat{M}_L P_L+\hat{M}_R P_R \right) \gamma^0 = \hat{M}^\dagger\,,
\end{equation} 
which is then inherited by the renormalized masses
\begin{equation}
  M_R=M_L^\dagger
\end{equation}
and the counterterms
\begin{equation}
  \label{eq:masspseudo}
  \delta M^{}_R=\delta M^\dagger_L\,.
\end{equation}
As the renormalized masses are diagonalized, the pseudo-hermiticity implies that the eigenvalues of the left and right masses must be the same, i.e.
\begin{equation}
  \label{eq:masstransform}
m\equiv(U_R^\dagger M_L^{} U_L^{})=(U_L^\dagger M_R^{} U_R^{})^\dagger\,.
\end{equation}
In turn, the diagonal mass matrix can be basis-transformed either as the left or the right mass matrix as needed.

The fields are renormalized as usual (although without square roots)
\begin{equation}
  \hat{\psi}_{u}=Z_{uu^\prime}\psi_{u^\prime}\,, \qquad
  \hat{\psi}_{d}=Z_{dd^\prime}\psi_{d^\prime}\,,
\end{equation}
with
\begin{equation}
  Z=I+\delta Z_L P_L+\delta Z_R P_R\,.
\end{equation}
It is not difficult to find that under basis transformations we have
\begin{equation}
\hat{\psi}^\prime=Z^\prime\psi^\prime=Z^\prime\mathcal{U}\psi=\mathcal{U}Z\psi\,,
\end{equation}
with 
\begin{equation}
  \label{eq:fieldtransform}
  Z=\mathcal{U}^\dagger Z^\prime \mathcal{U}=I+ U_L^\dagger Z^\prime_L U_L P_L
                                                +U_R^\dagger Z^\prime_R U_R P_R\,.
\end{equation}

Finally, upon going to the mass eigenstate basis, the $W$ vertex
\begin{align}
    \mathcal{L}_{Wud}=&-\frac{\hat{g}}{\sqrt{2}}\sum_{u,d} \overline{\hat{\psi}^\prime}_u \gamma^\mu \hat{W}^{+}_\mu \delta_{ud} P_L \hat{\psi}^\prime_d 
    \nonumber \\
  =&-\frac{\hat{g}}{\sqrt{2}}\sum_{u,d} \overline{\hat{\psi}}_u \gamma^\mu \hat{W}^{+}_\mu (U^\dagger_L D^{}_L)^{}_{ud} P_L \hat{\psi}_d
\end{align}
leads to the CKM matrix
\begin{equation}
  V_{ud}=(U^\dagger_L D^{}_L)^{}_{ud}\,,
\end{equation}
which is derived from renormalized parameters so that the counterterm is trivial by definition
\begin{equation}
\delta V_{ud}=0\,.  
\end{equation}

\subsubsection{Two point functions and renormalization conditions in the quark sector}

We may proceed to the self-energies in the quark sector, which are 
\begin{equation} 
i\Sigma_{ji}(\cancel{p})=
  \vcenter{\hbox{\begin{tikzpicture}
\begin{feynman}
  \vertex [blob] (m1) at (0,0){1PI};
  \vertex (a) at (-1.4,0.0) ;
  \vertex (b) at (1.4,0);
  \diagram*{
  (a)--[fermion, edge label=$i$](m1)--[fermion, edge label=$j$](b);
  };
\end{feynman}
\end{tikzpicture}}}
\hspace{1em}
+
\hspace{1em}
\vcenter{\hbox{\begin{tikzpicture}
\begin{feynman}
  \vertex [blob] (m1) at (0,0.75){1PI};
  \vertex (a) at (-1.4,0.0) ;
  \vertex (b) at (1.4,0);
  \vertex (c) at (0,0);
  \diagram*{
  (a)--[fermion, edge label=$i$](c)--[fermion, edge label=$j$](b);
  (c)--(m1);
  };
\end{feynman}
\end{tikzpicture}}}
\end{equation}
with the usual decomposition
\begin{equation}
\label{eq:selfdeco}
\begin{split}
      \Sigma_{ji}(\cancel{p})=
      &\Sigma^{\gamma L}_{ji}(p^2)\cancel{p} P_L+\Sigma^{\gamma R}_{ji}(p^2)\cancel{p}P_R
      +\Sigma^{sL}_{ji}(p^2)P_L+\Sigma^{sR}_{ji}(p^2)P_R\,.
\end{split}
\end{equation}
Here it is again assumed that the loop-functions are real and the self-energies obey the pseudo-hermiticity constraint
\begin{equation}
    \label{eq:selfpseudohermi}
    \begin{split}
        \left[ \Sigma^{\gamma L,\ \gamma R}_{ji}(p^2) \right]^\dagger=&\Sigma^{\gamma L,\ \gamma R}_{ji}(p^2)\,, 
        \\ 
        \left[ \Sigma^{sL,\ sR}_{ji}(p^2) \right]^\dagger=&\Sigma^{sR,\ sL}_{ji}(p^2)\,.  
    \end{split}
\end{equation}
By employing the definitions of the previous section, we have the following renormalized self-energy to first order
\begin{equation}
  \label{eq:renormalizedSigma}
  \Sigma_{ji}^R(\cancel{p})=\Sigma_{ji}(\cancel{p})+\gamma^0\delta Z_{ji}^\dagger \gamma^0\left( \cancel{p} - m_i\right)
  +\left( \cancel{p}-m_j \right)\delta Z_{ji} -\delta m_{ji}\,,
\end{equation}
where the indices $i,\ j$ are either $u,\ u^\prime$ or $d,\ d^\prime$.

The diagonal counterterms are fixed in the usual way. The requirement that the corrections vanish at the pole, $\Sigma^R_{ii}(\cancel{p})u_i=0$, produces the diagonal elements of the mass counterterm
\begin{align}
  \label{eq:massdiagct}
    \delta m_{ii}=&\frac{1}{2}\mathrm{Re}\Big[
                        m_i \Sigma^{\gamma R}_{ii}(m_i^2)
                        +\Sigma^{sL}_{ii}(m_i^2)
                        +m_i\Sigma^{\gamma L}_{ii}(m_i^2)
                        +\Sigma^{sR}_{ii}(m_i^2)\Big]\,.
\end{align}

The diagonal field counterterms come from the unit residue condition, $\lim_{\cancel{p}\rightarrow m_i}\frac{1}{\cancel{p}-m_i}\Sigma^{R}_{ii}(\cancel{p})u_i=u_i$, which results in 
\begin{align}
    \delta Z_{ii}=&\left( -\Sigma^{\gamma L}_{ii}(m_i^2)-\left.\partial_{p^2}\left( m^2_i\Sigma^{\gamma L}_{ii}(p^2)+m^2_i\Sigma^{\gamma R}_{ii}(p^2) +m_i\Sigma^{sL}_{ii}(p^2)+m_i\Sigma^{sR}_{ii}(p^2)\right)\right|_{p^2=m_i^2} \right)P_L 
    \nonumber \\ 
    &+\left( -\Sigma^{\gamma R}_{ii}(m_i^2)-\left.\partial_{p^2}\left( m^2_i\Sigma^{\gamma L}_{ii}(p^2)+m^2_i\Sigma^{\gamma R}_{ii}(p^2) +m_i\Sigma^{sL}_{ii}(p^2)+m_i\Sigma^{sR}_{ii}(p^2)\right)\right|_{p^2=m_i^2} \right)P_R\,.
\end{align}
For the off-diagonal elements we require to have a diagonal propagator, $\Sigma^R_{ji}(\cancel{p})u_i=0$, which leads to the following condition
\begin{equation}
\label{eq:fermionfieldmassrelation}
  \left[ (m_i^2-m_j^2)\delta Z_{ji}-m_j\delta m_{ji}-\delta m^\dagger_{ji} m_i \right]u_i=-(\cancel{p}+m_j)\Sigma^{}_{ji}(\cancel{p})u_i\,,\quad\text{for }i\neq j\,.
\end{equation}
Here we have two unknowns: the field and mass counterterms, but only one constraint. The anti-hermitian part of the above relation can be used to define the hermitian part of the field renormalization for $i\neq j$
\begin{align}
  \delta Z_{ji}^H=-\frac{1}{2}\frac{1}{m_i^2-m_j^2}&\left( 
      m^2_i \Sigma^{\gamma L}_{ji}(m_i^2)
    + m_i m_j  \Sigma^{\gamma R}_{ji}(m_i^2)
    +m_j \Sigma^{sL}_{ji}(m_i^2)
    +m_i \Sigma^{sR}_{ji}(m_i^2)-H.C.
    \right)P_L 
    \nonumber \\ \label{fermionfieldHerm}
    -\frac{1}{2}\frac{1}{m_i^2-m_j^2}&\left( 
      m^2_i \Sigma^{\gamma R}_{ji}(m_i^2)
    + m_i m_j  \Sigma^{\gamma L}_{ji}(m_i^2)
    +m_j \Sigma^{sR}_{ji}(m_i^2)
    +m_i \Sigma^{sL}_{ji}(m_i^2)-H.C.
    \right)P_R\,.
\end{align}
where $H.C.$ stands for Hermitian conjugation, which also acts on the arguments of the self-energy scalar functions, i.e. $[\Sigma(m_i^2)]^\dagger=\Sigma^\dagger(m_j^2)$. 

The hermitian part of Eq.~\eqref{eq:fermionfieldmassrelation} gives a relation between the anti-hermitian part of the field renormalization and the mass counterterms, however, unlike in the gauge sector, neither are fixed and one cannot solve for both the counterterms simultaneously. We go around the problem as in~\cite{drauksas2023} and define the anti-hermitian part of the field renormalization for $i\neq j$ as the coefficient of the $(m_i^2-m_j^2)$ mass structure
\begin{align}
  \delta Z^A_{ji}\equiv &-\frac{1}{2}\left.\left( 
    m^2_i \Sigma^{\gamma L}_{ji}(m_i^2)
    + m_i m_j \Sigma^{\gamma R}_{ji}(m_i^2)
    +m_j \Sigma^{sL}_{ji}(m_i^2)
    +m_i \Sigma^{sR}_{ji}(m_i^2)+H.C.
    \right)P_L\right|_{m_i^2-m_j^2} 
    \nonumber \\ \label{eq:fermionfieldoffdiag}
    &-\frac{1}{2}\left.\left( 
    m^2_i \Sigma^{\gamma R}_{ji}(m_i^2)
    + m_i m_j \Sigma^{\gamma L}_{ji}(m_i^2)
    +m_j \Sigma^{sR}_{ji}(m_i^2)
    +m_i \Sigma^{sL}_{ji}(m_i^2)+H.C.
    \right)P_R\right|_{m_i^2-m_j^2}\,.
\end{align}
This definition is motivated by the Nielsen Identities~\cite{nielsen1975,gambino2000} and the mass structures, see~\cite{drauksas2023} and references therein. The practical implementation of these definitions is discussed in the next section, while here we give the simple result for the Standard Model quarks that we get from Eq.~\eqref{eq:ZAfullsimple} (or the SM-specific version Eq.~\eqref{eq:simpleZA}) of the next section
\begin{equation}
\label{eq:quarkZAsmresult}
\begin{split}
  \delta Z^A_{uu^\prime}&=-\frac{\alpha}{32\pi s^2 m_W^2}\sum_{d} V^{}_{ud} V^\star_{u^\prime d}\left( m_d^2-m_{u^\prime}^2+\xi_W m_W^2 \right)B_0\left[ m_{u^\prime}^2, m_d^2, \xi_W m_W^2 \right]P_L - H.C.\,,\\
  \delta Z^A_{dd^\prime}&=-\frac{\alpha}{32\pi s^2 m_W^2}\sum_{u} V^{}_{ud^\prime} V^\star_{u d}\left( m_u^2-m_{d^\prime}^2+\xi_W m_W^2 \right)B_0\left[ m_{d^\prime}^2, m_u^2, \xi_W m_W^2 \right]P_L - H.C.\,,
\end{split}
\end{equation}
where $\alpha=e^2/(4\pi)$ is the fine structure constant and $B_0\left[ \dots \right]$ is the Passarino--Veltman (PV) function~\cite{thooft1979,passarino1979} with the conventions of~\cite{denner2006}. It is worth noting that the resulting anti-hermitian parts of the field renormalization are purely left-handed, reflecting the chirality of the weak interaction, and minimal in the sense that they consist only of gauge-dependent terms needed to cancel the gauge-dependence in the $W$ amplitude. In addition, one may explicitly check that the anti-hermitian parts are UV finite in the SM, which means that these counterterms are trivial in $\overline{\mathrm{MS}}$ schemes.

Having fully defined the field renormalization we may solve for the mass counterterms, which are
\begin{align}
  \delta m_{ji}=&\frac{1}{2}\left(
  m_i \Sigma^{\gamma R}_{ji}(m_i^2)
  +\Sigma^{sL}_{ji}(m_i^2)
  +\left[m_i \Sigma^{\gamma L}_{ji}(m_i^2) 
          + \Sigma^{sR}_{ji}(m_i^2) \right]^\dagger \right)P_L
  \nonumber \\ 
  &+\frac{1}{2}\left(
  m_i \Sigma^{\gamma L}_{ji}(m_i^2)
  +\Sigma^{sR}_{ji}(m_i^2)
  +\left[ m_i \Sigma^{\gamma R}_{ji}(m_i^2)
          + \Sigma^{sL}_{ji}(m_i^2) \right]^\dagger \right)P_R
  \nonumber \\ \label{eq:massct}
    &-m_j \delta Z^A_{ji}+m_i \gamma^0\delta Z^A_{ji}\gamma^0\,, \qquad \text{for } i\neq j\,. 
\end{align}

Note that the real part of this reproduces the diagonal ($i=j$) case, but is otherwise different. With these mass and field counterterm definitions the CKM counterterm is trivial
\begin{eqnarray}
  \delta V^{}_{ud}=0\,.
\end{eqnarray}
Note that the counterterms defined in our scheme~\cite{drauksas2023} presented above comprise a fully consistent On-Shell scheme without the need of any additional counterterms, as is not the case usually~\cite{gambino1999,yamada2001,dienerkniehl,liao2004,kniehl2006,kniehl2009a}. 

Note that the l.h.s. of Eq.~\eqref{eq:fermionfieldmassrelation} implies that the field and mass counterterms are degenerate, which also implies that they are degenerate with mixing angle counterterms via Eq.~\eqref{eq:degeneracy} --- this is a problem since the mass counterterm has a distinct physical meaning. On the other hand, it is not uncommon to use this apparent degeneracy and define the mixing matrix counterterms through the mass counterterms, e.g.~\cite{kniehl2006}. However, we believe this to be a misconception that arises from not considering the mass structures. Using the mass structures the degeneracy between mass and field (and mixing matrix) counterterms is broken, while the degeneracy between field and mixing matrix counterterms implied by Eq.~\eqref{eq:degeneracy} remains. Therefore, it is expedient to further investigate the mass structures.

\subsection{Fermion field renormalization in practice}
\label{sec:method}

Here we provide a practical prescription for the definition of the anti-hermitian part of the field renormalization in Eq.~\eqref{eq:fermionfieldoffdiag} and getting the results of Eq.~\eqref{eq:quarkZAsmresult}. The definition and the prescription rely on the presence of the distinct mass structure of ${m_i^2-m_j^2}$, which is not trivially seen on the r.h.s. of Eq.~\eqref{eq:fermionfieldmassrelation}. The outline of the path to the prescription is as follows: 
\begin{itemize}
  \item we propose a fine-grained decomposition of the fermion self-energy in terms of the external masses in Eq.~\eqref{eq:selfextradeco} and list the needed properties;
  \item we investigate Eq.~\eqref{eq:fermionfieldmassrelation} in terms of the new decomposition and identify the terms proportional to $m_j$ as candidate contributions to the mass counterterms, then the remaining terms are proportional to $m_i^2-m_j^2$ and are possible contributions to the anti-hermitian part of the field renormalization;
  \item in Appendix~\ref{sec:AnielsenSE} we have provided the Nielsen identities, with which we check that the counterterm candidates have the needed gauge-dependence;
  \item by the basis transformation properties of the extended decomposition we identify UV finite or divergent off-diagonal self-energy scalar functions and notice, that the candidates for anti-hermitian part of the field renormalization include incorrect UV divergences, which therefore, must be subtracted;
  \item we further find that the needed UV subtraction is gauge-independent and can be performed without altering gauge-dependence properties;
  \item we provide the final result in Eq.~\eqref{eq:ZAfullsimple}.
\end{itemize}

\subsubsection{The fine-grained decomposition of the fermion self-energy}

First, it is notable that the situations for the hermitian and anti-hermitian parts of the field renormalization are different. In the hermitian part of the field renormalization defined in Eq.~\eqref{fermionfieldHerm} the factor of $1/(m_i^2-m_j^2)$ cancels once the self-energies are Taylor expanded around one of the masses. In contrast, such an expansion for the anti-hermitian part does not produce the needed mass structure, therefore it must appear as some intrinsic combination of the self-energies, in other words the self-energies are likely not fine-grained enough to be sensitive to this structure and a different decomposition than in Eq.~\eqref{eq:selfdeco} is needed.

As we are after the masses it is worth to consider a decomposition which highlights the dependence on the masses. One can separate the internal masses, which are summed over in the loop, from the external masses, which have their indices fixed to either $i$ or $j$, with the help of mixing matrices. The 2-point function has only two external legs and if the coupling to the external leg is proportional to the mixing matrix, then it forbids the external indices to from appearing inside the diagram. On the other hand, if the coupling is diagonal, then the internal summation indices become fixed to the external ones. Terms in which the external indices are not propagated to the loop have a simple structure in terms of the masses $m_i$ and $m_j$. One simply has to notice that in 2-point functions only two couplings can be coupled to the external legs, and in turn such terms are at most bilinear in the external fermion masses, equivalently, to the Yukawa couplings in spontaneously broken models. Taking this into account, we use the symbol $\times$ to denote that there is no mixing matrix coupling directly to the external leg, the symbol $\circ$ to denote that there is a mixing matrix, but there is no external mass, and a $\bullet$ for the case with the mixing matrix and first power of the corresponding external mass. Therefore, at 1-loop, each scalar function in the usual self-energy decomposition in Eq.~\eqref{eq:selfdeco} may be further decomposed into these scalar functions
\begin{equation}
\label{eq:selfextradeco}
\begin{split}
      \Sigma^{X}_{ji}(p^2)=&
                  \Sigma^{X,\times\times}_{ji}(p^2)
                  +\Sigma^{X,\circ\circ}_{ji}(p^2) 
                  +m_i \Sigma^{X,\circ\bullet}_{ji}(p^2)
                  + m_j \Sigma^{X,\bullet\circ}_{ji}(p^2)
                  + m_i m_j \Sigma^{X,\bullet\bullet}_{ji}(p^2)\,,\
\end{split}
\end{equation}
where $X=\{\gamma L,\ \gamma R,\ sL,\ sR \}$. Beyond 1-loop one may have mixed contributions, where one of the external legs is directly coupled through the mixing matrix, while the other is not. The symbol $\times$ effectively denotes that the external index has propagated to the internal loop masses and there is a more complex dependence on the external masses. On the other hand, functions without $\times$ in their superscript are completely free of the external fermion masses. Another simplification is that at 1-loop $\Sigma^{X, \times\times}$ is fully diagonal. 

A subtlety comes from the tadpole diagrams since there is only one coupling for both of the external legs so that the superscripts should only contain one of the symbols $\times,\circ$ or $\bullet$. However, if the coupling to the external legs is non-diagonal we simply use the masses to distribute the contributions to $\Sigma^{X, \circ\circ}$, $\Sigma^{X, \circ\bullet}$ or $\Sigma^{X, \bullet\circ}$ functions as the tadpoles are at most linear in the external masses and cannot contribute to $\Sigma^{X, \bullet\bullet}$. Otherwise, if the coupling is diagonal, the tadpoles contribute to $\Sigma^{X,\times\times}$. 

Note that the loop integrals should be written in terms of basis integrals and the unitarity of the mixing matrix should be used before doing the decomposition. For a concrete example, the SM up-type quark self-energy at one loop contains the following contribution (up to irrelevant factors) $m_u m_{u^\prime}\sum_d V_{ud}V^{\star}_{u^\prime d} B_1[p^2,m^2_d,\xi_W m_W^2]$, but the PV function may be reduced into $A_0$ and $B_0$ functions. One of the terms produced by the reduction is $m_u m_{u^\prime}\sum_d V_{ud}V^{\star}_{u^\prime d} A_0(\xi_W m_W^2)=m_u^2\delta_{uu^\prime }A_0(\xi_W m_W^2)$, which then belongs to $\Sigma^{X,\times\times}$ instead of $\Sigma^{X, \bullet\bullet}$. 

Continuing to the properties of the decomposition, we note that since it holds for arbitrary momenta, one may use pseudo-hermiticity in Eq.~\eqref{eq:selfpseudohermi} to further constrain the scalar-functions
\begin{align}
  \Sigma^{\gamma L, \gamma R,\times\times, \circ\circ}_{ji}(p^2)&=\left[ \Sigma^{\gamma L, \gamma R,\times\times,\circ\circ}_{ji}(p^2) \right]^\dagger \,,
  \nonumber \\ 
  \Sigma^{\gamma L, \gamma R,\bullet\circ}_{ji}(p^2)&=\left[ \Sigma^{\gamma L, \gamma R,\circ\bullet}_{ji}(p^2) \right]^\dagger \,,
  \nonumber \\ 
  \Sigma^{\gamma L, \gamma R,\bullet\bullet}_{ji}(p^2)&=\left[ \Sigma^{\gamma L, \gamma R,\bullet\bullet}_{ji}(p^2) \right]^\dagger \,,
  \nonumber \\
  \Sigma^{sL,sR,\times\times, \circ\circ}_{ji}(p^2)&= \left[ \Sigma^{sR,sL,\times\times,\circ\circ}_{ji}(p^2) \right]^\dagger\,,
  \nonumber \\ 
  \Sigma^{sL,sR,\circ\bullet}_{ji}(p^2)&= \left[ \Sigma^{sR,sL,\bullet\circ}_{ji}(p^2) \right]^\dagger\,,
  \nonumber \\
  \Sigma^{sL,sR,\bullet\bullet}_{ji}(p^2)&= \left[ \Sigma^{sR,sL,\bullet\bullet}_{ji}(p^2) \right]^\dagger
  \,.
\end{align}
Just like pseudo-hermiticity itself, these restrictions hold below particle production thresholds or if the absorptive parts are dropped. 

The scalar functions are also endowed with certain basis transformation properties. Taking the basis transformation of the up-type quarks in Eq.~\eqref{eq:fermionrotation} as an example, we see that already in the usual decomposition in Eq.~\eqref{eq:selfdeco}, the self-energies fall into the basis transformation classes implied by Eqs.~\eqref{eq:masstransform} and~\eqref{eq:fieldtransform},
\begin{align}
  \text{left field-like: }\quad\Sigma^{\gamma L}=&U_L^\dagger(\Sigma^{\gamma L})^\prime U_L\,, 
  \nonumber \\
  \text{right field-like: }\quad\Sigma^{\gamma R}=&U_R^\dagger (\Sigma^{\gamma R})^\prime U_R\,, 
  \nonumber \\
  \text{left mass-like: }\quad\Sigma^{sL}=&U_R^\dagger(\Sigma^{sL})^\prime U_L\,, 
  \nonumber \\
  \text{right mass-like: }\quad\Sigma^{sR}=&U_L^\dagger(\Sigma^{sR})^\prime U_R\,.
\end{align}
Let us take $\Sigma^{\gamma L}$ as an example and derive how the scalar functions of the fine-grained decomposition in Eq.~\eqref{eq:selfextradeco} transform. First, we invert the transformation
\begin{equation}
\begin{split}
 (\Sigma^{\gamma L})^\prime_{kl}=&\sum_{j,i}(U_L)_{kj}\Sigma^{\gamma L}_{ji}(U_L^\dagger)_{il}\\
    =&\sum_{j,i}(U_L)_{kj}\left(\Sigma^{\gamma L,\times\times}+\Sigma^{\gamma L,\circ\circ}+ \Sigma^{\gamma L,\circ\bullet}_{ji}m_i+m_j\Sigma^{\gamma L,\bullet\circ}_{ji}+m_j\Sigma^{\gamma L,\bullet\bullet}_{ji}m_i\right)(U_L^\dagger)_{il}\\
    =&
    (U_L\Sigma^{\gamma L, \times\times}U_L^\dagger)_{kl}\\ 
    &+(U_L\Sigma^{\gamma L, \circ\circ}U_L^\dagger)_{kl}
    +(U_L\Sigma^{\gamma L, \circ\bullet}mU_L^\dagger)_{kl}
    +(U_L m \Sigma^{\gamma L, \bullet\circ}U_L^\dagger)_{kl}
    +(U_Lm\Sigma^{\gamma L, \bullet\bullet}mU_L^\dagger)_{kl}\,.
\end{split}
\end{equation}
Next, we comply with the mass-like basis transformations of Eq.~\eqref{eq:masstransform} and insert identity matrices expressed as $I=U^\dagger_R U_R^{}=U_R^{} U_R^\dagger$ to consistently produce left or right masses
\begin{equation}
\begin{split}
  (\Sigma^{\gamma L})^\prime=&
        U_L\Sigma^{\gamma L, \times\times}U_L^\dagger\\
        &+U_L\Sigma^{\gamma L, \circ\circ}U_L^\dagger
        +\left[U_L\Sigma^{\gamma L, \circ\bullet}U_R^\dagger\right]M_L
        +M_R \left[U_R\Sigma^{\gamma L, \bullet\circ}U_L^\dagger\right]
        +M_R \left[U_R\Sigma^{\gamma L, \bullet\bullet}U_R^\dagger\right] M_L \,.
\end{split}
\end{equation}
Here it is simple to identify the (inverse) basis transformation class of each scalar function. One can apply an analogous procedure to all the scalar functions and find that they belong to the 4 classes as follows
\begin{align}
  \text{left field-like: }\quad\{\Sigma^{\gamma L,\times\times},
      \Sigma^{\gamma L,\circ\circ}, \Sigma^{\gamma R,\bullet\bullet}, 
      \Sigma^{sL,\bullet\circ}, \Sigma^{sR,\circ\bullet}\}\,, 
  \nonumber \\
  \text{right field-like: }\quad \{\Sigma^{\gamma R,\times\times},
      \Sigma^{\gamma R,\circ\circ}, \Sigma^{\gamma L,\bullet\bullet},
      \Sigma^{sR,\bullet\circ}, \Sigma^{sL,\circ\bullet} \}\,, 
  \nonumber \\
  \text{left mass-like: }\quad \{ \Sigma^{sL,\times\times},
      \Sigma^{sL,\circ\circ}, \Sigma^{sR,\bullet\bullet},
      \Sigma^{\gamma L,\bullet\circ}, \Sigma^{\gamma R,\circ\bullet}\}\,, 
  \nonumber \\
  \text{right mass-like: }\quad \{\Sigma^{sR,\times\times},
      \Sigma^{sR,\circ\circ}, \Sigma^{sL,\bullet\bullet},
      \Sigma^{\gamma R,\bullet\circ}, \Sigma^{\gamma L,\circ\bullet}\}\,.
  \label{eq:extrabasis}
\end{align}

Finally, to conclude the introduction of the fine-grained decomposition, we note that the tree-level self-energy (inverse propagator) does not fall into the decomposition as there is no separation between the two external legs.  

\subsubsection{Selecting the counterterm candidates}

We return to Eq.~\eqref{eq:fermionfieldmassrelation} and have the decompositions in Eqs.~\eqref{eq:selfdeco} and~\eqref{eq:selfextradeco} in mind. Importantly, we have to consider only off-diagonal elements, which means that the self-energy functions $\Sigma^{X,\times\times}$ do not contribute. Without yet considering hermitian conjugation, the appearance of the masses $m_i^2$ and $m_j^2$ is not symmetric, since acting with the spinor $u_i$ can only produce the mass $m_i$. Therefore, on the r.h.s. $m_j$ can come either from the factor $\cancel{p}+m_j$ or from the decomposition in Eq.~\eqref{eq:selfextradeco}, therefore, the resulting expression is at most quadratic in $m_j$. On the l.h.s., the terms linear in $m_j$ multiply $\delta m_{Lji}$ in the left-handed part and $\delta m^\dagger_{Lji}$ in the right-handed part. Considering the coefficients of $m_j$ on the r.h.s. as contributions to the mass counterterms, we find the following

\begin{align}
  \delta m_{Lji}\supset & \left[\Sigma^{sL,\circ\circ}_{ji}(m_i^2)
                            +m_i^2\left(
                                \Sigma^{\gamma L,\bullet\circ}_{ji}(m_i^2) 
                                +\Sigma^{\gamma R,\circ\bullet}_{ji}(m_i^2)
                                +\Sigma^{sR,\bullet\bullet}_{ji}(m_i^2) 
                              \right)\right]
                        \nonumber \\ 
                        &+ m_i\left[\Sigma^{\gamma R,\circ\circ}_{ji}(m_i^2)
                                +\Sigma^{sL,\circ\bullet}_{ji}(m_i^2)
                                +\Sigma^{sR,\bullet\circ}_{ji}(m_i^2)
                                + m_i^2 \Sigma^{\gamma L,\bullet\bullet}_{ji}(m_i^2)\right]\,, \quad\text{for } i\neq j\,,
\label{eq:dmla}
\end{align}
and
\begin{align}
  \delta m^\dagger_{Lji}\supset & \left[\Sigma^{sR,\circ\circ}_{ji}(m_i^2)               
                      +m_i^2\left(
                                \Sigma^{\gamma R,\bullet\circ}_{ji}(m_i^2) 
                                +\Sigma^{\gamma L,\circ\bullet}_{ji}(m_i^2)
                                +\Sigma^{sL,\bullet\bullet}_{ji}(m_i^2) 
                              \right)\right]
                        \nonumber \\ 
                        &+m_i\left[\Sigma^{\gamma L,\circ\circ}_{ji}(m_i^2)
                                +\Sigma^{sR,\circ\bullet}_{ji}(m_i^2)
                                +\Sigma^{sL,\bullet\circ}_{ji}(m_i^2)+m_i^2\Sigma^{\gamma R,\bullet\bullet}_{ji}(m_i^2) \right]\,,\quad\text{for } i\neq j\,.
  \label{eq:dmlb}
\end{align}
Here the pseudo-hermiticity of Eq.~\eqref{eq:masspseudo} does not hold, but this can be easily remedied by adding the hermitian conjugate of Eq.~\eqref{eq:dmlb} to Eq.~\eqref{eq:dmla} and vice versa, but this is not really needed. By using the Nielsen identities and Eqs.~\eqref{eq:treesigmalambda1}--\eqref{eq:treesigmalambda36} from Appendix~\ref{sec:AnielsenSE} it is simple to find that these contributions to the mass counterterms are indeed gauge-independent at 1-loop. Even more so, the contributions in the square brackets are gauge-independent separately and even for arbitrary momenta and indices $i$ and $j$, i.e.
\begin{equation}
  \label{eq:firstlinedmla}
  \partial_\xi\left(\Sigma^{sL,\circ\circ}_{ji}(p^2)
                            +p^2\left(
                                \Sigma^{\gamma L,\bullet\circ}_{ji}(p^2) 
                                +\Sigma^{\gamma R,\circ\bullet}_{ji}(p^2)
                                +\Sigma^{sR,\bullet\bullet}_{ji}(p^2) 
                              \right) \right)=0\,,
\end{equation}
\begin{equation}
  \label{eq:firstlinedmlb}
  \partial_\xi\left(\Sigma^{sR,\circ\circ}_{ji}(p^2)               
                      +p^2\left(
                                \Sigma^{\gamma R,\bullet\circ}_{ji}(p^2) 
                                +\Sigma^{\gamma L,\circ\bullet}_{ji}(p^2)
                                +\Sigma^{sL,\bullet\bullet}_{ji}(p^2) 
                              \right) \right)=0\,,
\end{equation}
\begin{equation}
  \label{eq:secondlinedmla}
  \partial_\xi\left(\Sigma^{\gamma R,\circ\circ}_{ji}(p^2)
                                +\Sigma^{sL,\circ\bullet}_{ji}(p^2)
                                +\Sigma^{sR,\bullet\circ}_{ji}(p^2)
                                + p^2 \Sigma^{\gamma L,\bullet\bullet}_{ji}(p^2)  \right)=0\,,
\end{equation}
and
\begin{equation}
  \label{eq:secondlinedmlb}
  \partial_\xi \left( \Sigma^{\gamma L,\circ\circ}_{ji}(p^2)
                                +\Sigma^{sR,\circ\bullet}_{ji}(p^2)
                                +\Sigma^{sL,\bullet\circ}_{ji}(p^2)+p^2\Sigma^{\gamma R,\bullet\bullet}_{ji}(p^2) \right)=0\,.
\end{equation}

Consulting Eq.~\eqref{eq:extrabasis} we notice that Eq.~\eqref{eq:firstlinedmla} and Eq.~\eqref{eq:firstlinedmlb} transform like the masses under basis transformations and are hermitian conjugates of each other, while Eq.~\eqref{eq:secondlinedmla} and Eq.~\eqref{eq:secondlinedmlb} transform like the fields and are hermitian. The mass-like terms also present in the first lines of Eq.~\eqref{eq:dmla} and Eq.~\eqref{eq:dmlb} can be considered as genuine corrections to the mass matrix. In the mass eigenstate basis the renormalized mass is diagonal, therefore the mass-like contributions must consist of UV finite\footnote{A similar argument can be made in seesaw models, e.g. see Ref.~\cite{grimus2002}.} scalar functions for $i \neq j$. For example, for $m_i=0$ one gets the usual pieces of $\Sigma^{sL,\circ\circ}(0)$ and $\Sigma^{sR,\circ\circ}(0)$ associated with the radiative masses, which is UV finite. On the other hand, the field-like terms present on the second lines of Eq.~\eqref{eq:dmla} and Eq.~\eqref{eq:dmlb} are not limited by the diagonality of the mass matrix and are gauge-independent, but possibly UV divergent contributions to the mass counterterm. In the mass counterterm the correct basis transformations of these field-like terms are ensured by the multiplication by the mass $m_i$, which can transform either as the left or right mass as needed.

For completeness, the diagonal mass counterterm in Eq.~\eqref{eq:massdiagct} implies the following gauge-independent combination of self-energy scalar functions at 1-loop
\begin{align}
  \partial_\xi\Big(m_i \Sigma^{\gamma R,\times\times}_{ii}(m_i^2)
                        +\Sigma^{sL,\times\times}_{ii}(m_i^2)
                        +m_i\Sigma^{\gamma L,\times\times}_{ii}(m_i^2)
                        +\Sigma^{sR,\times\times}_{ii}(m_i^2)\Big)=0\,.
\end{align}
However, here the gauge-dependence vanishes only on-shell, $p^2=m_i^2$, and is otherwise proportional to $p^2-m_i^2$, as can be checked by using Eqs.~\eqref{eq:treesigmalambda1}--\eqref{eq:treesigmalambda36}.

Upon inserting the contributions of Eq.~\eqref{eq:dmlb} and Eq.~\eqref{eq:dmla} as if they are the full mass counterterms in Eq.~\eqref{eq:fermionfieldmassrelation}, acting with and then dropping the spinors, and taking the hermitian part, we find that all the remaining terms on the r.h.s. are proportional to $m_i^2-m_j^2$, which is exactly what is needed to apply Eq.~\eqref{eq:fermionfieldoffdiag}. By again using the Nielsen identities and Eqs.~\eqref{eq:treesigmalambda1}--\eqref{eq:treesigmalambda36} from Appendix~\ref{sec:AnielsenSE}, one can check the gauge-dependence of these remaining terms at 1-loop and get

\begin{align}
    &\partial_\xi \left(m_i \Sigma^{\gamma R,\bullet\circ}_{ji}(m_i^2)
    +m_i^2 \Sigma^{\gamma R,\bullet\bullet}_{ji}(m_i^2)
    +\Sigma^{sL,\bullet\circ}_{ji}(m_i^2)
    + m_i \Sigma^{sL,\bullet\bullet}_{ji}(m_i^2)\right)P_L
    \nonumber \\
    =&-\left(m_i \bar{\Lambda}^{\gamma R}_{ji}(m_i^2)+\bar{\Lambda}^{sL}_{ji}(m_i^2)\right)P_L\,,
    \label{eq:ZLgauge} \\
    &\partial_\xi \left(  m_i \Sigma^{\gamma L,\bullet\circ}_{ji}(m_i^2)
    +m_i^2 \Sigma^{\gamma L,\bullet\bullet}_{ji}(m_i^2)
    +\Sigma^{sR,\bullet\circ}_{ji}(m_i^2)
    + m_i \Sigma^{sR,\bullet\bullet}_{ji}(m_i^2) \right)P_R
    \nonumber \\
    =&-\left(m_i \bar{\Lambda}^{\gamma L}_{ji}(m_i^2)+\bar{\Lambda}^{sR}_{ji}(m_i^2)\right)P_R\,,
    \label{eq:ZRgauge}
\end{align}
which is known to contain all the gauge-dependence in the field renormalization~\cite{espriu2002,yamada2001,drauksas2023} in the On-Shell scheme.  

Naively, one could take the anti-hermitian part of combinations of the self-energy scalar functions on the l.h.s. of Eqs.~\eqref{eq:ZLgauge} and~\eqref{eq:ZRgauge} and use these as the anti-hermitian parts of the field renormalization. However, while the decomposition in Eq.~\eqref{eq:selfextradeco} allows to pick the gauge-dependent parts, it is still not fine-grained enough to correctly separate the UV parts. For example, taking just the UV parts and using pseudo-hermiticity we may see the contributions of the possibly UV divergent scalar self-energies of Eqs.~\eqref{eq:ZLgauge} and~\eqref{eq:ZRgauge} in Eq.~\eqref{eq:fermionfieldoffdiag}

    \begin{align}
      \delta Z^A_{ji}\supset &-\Bigg( 
        m_i^2m_j^2 \Sigma^{\gamma R,\bullet\bullet}_{ji}
        +(m_i^2+m_j^2) \Sigma^{sL,\bullet\circ,H}_{ji}
        -(m_i^2-m_j^2) \Sigma^{sL,\bullet\circ,A}_{ji}
        \Bigg)P_L\Bigg|_{m_i^2-m_j^2, \text{ UV}}
        \nonumber \\
        &
        -\Bigg( 
        m_i^2m_j^2 \Sigma^{\gamma L,\bullet\bullet,}_{ji}
        +(m_i^2+m_j^2) \Sigma^{sR,\bullet\circ,H}_{ji}
        -(m_i^2-m_j^2) \Sigma^{sR,\bullet\circ,A}_{ji}
        \Bigg)P_R \Bigg|_{m_i^2-m_j^2, \text{ UV}}
        \nonumber \\
        =&\left( \Sigma^{sL,\bullet\circ,A}_{ji} P_L+ \Sigma^{sR,\bullet\circ,A}_{ji}P_R \right)\Big|_{\text{UV}}\,,
        \label{eq:dZAUVparts}
    \end{align}
where we have dropped the momentum dependence, which is irrelevant for the UV parts, $\Sigma^{X,\bullet\circ,H}$ is the hermitian and $\Sigma^{X,\bullet\circ,A}$ is the anti-hermitian part of the respective self-energies. We see that there may be UV divergent contributions to the anti-hermitian part of the field renormalization only from $\Sigma^{sL,\bullet\circ}$ or $\Sigma^{sR,\bullet\circ}$\footnote{This is an update to our previous statement that the anti-hermitian part must be UV finite in~\cite{drauksas2023}. In this work some UV divergences are now allowed by the fine-grained decomposition in Eq.~\eqref{eq:selfextradeco}}. On the other hand, the UV divergences of the other functions in Eqs.~\eqref{eq:ZLgauge} and~\eqref{eq:ZRgauge} must not be attributed to the anti-hermitian part of the field renormalization. To have this ability we note that the UV divergences do not depend on the 'details' of the loop, so we may rewrite the self-energies as follows
\begin{align}
  \Sigma^{X,Y}_{ji}(p^2)=&\tilde{\Sigma}^{X,Y}_{ji}(p^2) +\left[ \Sigma^{X,Y}_{ji}(p^2)-\tilde{\Sigma}^{X,Y}_{ji}(p^2) \right]
  \nonumber \\
                  =&\tilde{\Sigma}^{X,Y}_{ji}(p^2) + \overline{\Sigma}^{X,Y}_{ji}(p^2)\,,
\end{align}
where all the masses in the loop integrals are taken to 0 and the absorptive parts of loop functions are dropped in $\tilde{\Sigma}^{X,Y}_{ji}(p^2)$ with  $Y=\{\times\times,\circ\circ,\circ\bullet,\bullet\circ,\bullet\bullet\}$. For example, a loop propagator like $\frac{1}{k^2-m^2}$ becomes $\frac{1}{k^2}$ in $\tilde{\Sigma}$. Note however, that such a replacement singles out the UV parts\footnote{Note that this puts the UV divergences from tadpoles to 0 in dimensional regularization, but this is irrelevant for our final result in Eq.~\eqref{eq:ZAfullsimple} as there we use the subtraction only for functions, which cannot have tadpole contributions.}, but it also selects UV finite momentum-dependent terms like $\log(p^2)$, which do not obey pseudo-hermiticity on-shell. 

Briefly returning to Eq.~\eqref{eq:dZAUVparts} we remind that the gauge-dependent parts must be multiplied by $m_i^2-m_j^2$ as instructed by the Nielsen identities in Eqs.~\eqref{eq:ZLgauge} and~\eqref{eq:ZRgauge}, where this mass structure cancels out. In turn, the UV parts of $\Sigma^{\gamma L,\bullet\bullet}_{ji}$ and $\Sigma^{\gamma R,\bullet\bullet}_{ji}$ must be gauge-independent as they are multiplied by $m_i^2 m_j^2$ instead. The same conclusion follows by noticing that the UV finiteness of the first lines of Eq.~\eqref{eq:dmla} and Eq.~\eqref{eq:dmlb} combined with the Nielsen identities in Eqs.~\eqref{eq:treesigmalambda1}--\eqref{eq:treesigmalambda36} show that the only UV divergent functions of the decomposition in Eq.~\eqref{eq:nielsenextra} are $\Lambda^{sL,\circ\circ}_{ji}$, $\Lambda^{sR,\circ\circ}_{ji}$, $\bar{\Lambda}^{sL,\circ\circ}_{ji}$, and $\bar{\Lambda}^{sR,\circ\circ}_{ji}$. Then one easily deduces that at 1-loop the gauge-dependent parts of $\Sigma^{\gamma R,\bullet\bullet}_{ji}$ and $\Sigma^{\gamma L,\bullet\bullet}_{ji}$ are UV finite, which is important for our main result in Eq.~\eqref{eq:ZAfullsimple}.

These are the last ingredients needed to promote the candidates for the anti-hermitian part of the field renormalization in Eqs.~\eqref{eq:ZLgauge} and~\eqref{eq:ZRgauge} into proper counterterms.

\subsubsection{Prescription for the anti-hermitian part of the field renormalization}
\label{sec:themainresult}

We propose the following practical formula for the anti-hermitian part of the field renormalization for $i\neq j$ at 1-loop
  \begin{equation}
    \label{eq:ZAfullsimple}
    \begin{split}
      \delta Z^A_{ji}=&
      \frac{1}{2}\left( m_i \Sigma^{\gamma R,\bullet\circ}_{ji}(m_i^2)
            +m_i^2 \overline{\Sigma}^{\gamma R,\bullet\bullet}_{ji}(m_i^2)
            +\Sigma^{sL,\bullet\circ}_{ji}(m_i^2)
            + m_i \Sigma^{sL,\bullet\bullet}_{ji}(m_i^2)-H.C.\right) P_L \\
      &+\frac{1}{2}\left( m_i \Sigma^{\gamma L,\bullet\circ}_{ji}(m_i^2)
              +m_i^2 \overline{\Sigma}^{\gamma L,\bullet\bullet}_{ji}(m_i^2)
              +\Sigma^{sR,\bullet\circ}_{ji}(m_i^2)
              + m_i \Sigma^{sR,\bullet\bullet}_{ji}(m_i^2)-H.C.\right) P_R
              \,.
    \end{split}
  \end{equation}
Importantly, note that by using the subtracted self-energy scalar functions $\overline{\Sigma}^{\gamma L,\bullet\bullet}_{ji}$ and $\overline{\Sigma}^{\gamma R,\bullet\bullet}_{ji}$ we did not alter the gauge-dependence because the subtracted UV divergences are gauge-independent as discussed right above the start of this section. This would not be the case for the function $\Sigma^{sL,\bullet\circ}_{ji}$ or $\Sigma^{sR,\circ\bullet}_{ji}$, but they are left not subtracted.

We consider to have found a model-independent prescription for $\delta Z_{ji}^A$ such that the mixing matrix counterterms are trivial. Note, that this produces a field renormalization counterterm that contains all the relevant gauge-dependence since all the gauge-dependence is necessarily multiplied by $m_i^2-m_j^2$. To get the full mass counterterm it is convenient to insert $\delta Z^A$ of Eq.~\eqref{eq:ZAfullsimple} into Eq.~\eqref{eq:massct}. 

It is important to note that to find $\delta Z^A$ we needed the fine-grained decomposition in Eq.~\eqref{eq:selfextradeco} and to split off the UV parts in some convenient way, for which we just neglected the loop masses. We have constructed $\delta Z^A$ to 1-loop order since we consider a 1-loop decay in the following sections. For the possibility of extension to higher orders see Appendix~\ref{secA:extensions}.

In the Standard Model Eq.~\eqref{eq:ZAfullsimple} simplifies since most of the scalar functions simply vanish for $i\neq j$ and, in addition, the subtraction is not even needed, therefore we have 
\begin{equation}
\label{eq:simpleZA}
\begin{split}
    \mathrm{SM:}\quad \delta Z^A_{ji}=\frac{1}{2}\Bigl(&m_i^2 \Sigma^{\gamma R,\bullet\bullet}_{ji}(m_i^2) +\Sigma^{sL,\bullet\circ}_{ji}(m_i^2)- H.C.\Bigr) P_L\,,\qquad \text{for } i\neq j\,.
\end{split}
\end{equation}

For an example that goes beyond the Standard model, we have tested this prescription for the Grimus--Neufeld model~\cite{GrimusNeufeld}, that has a Majorana singlet in addition to the 3 SM neutrinos, and the mixing matrix is split into 2 parts
  $U=\begin{pmatrix}
    U_L^\dagger & U_R^T
  \end{pmatrix}$
where $U_L$ is $3\times 4$ and $U_R$ is $1\times 4$. Due to the construction of the model, instead of the full mixing matrix $U$, either $U_L$ or $U_R$, which are not unitary by themselves, appear in the Lagrangian separately. Even with this contrived implementation of mixing matrices, the formula for $\delta Z^A$ in Eq.~\eqref{eq:ZAfullsimple} still works, although one must use the subtracted self-energy scalar functions, otherwise the UV divergences are distributed to the wrong counterterms due to non-unitarity of the non-full mixing matrix.

Even though one uses the subtracted parts of the self-energy scalar functions in Eq.~\eqref{eq:ZAfullsimple}, this is only a convenient way of separating the UV parts, but the self-energies are really evaluated only on-shell at $p^2=m_i^2$ or $p^2=m_j^2$, so that the scheme is indeed a genuine OS scheme based on the traditional On-Shell conditions. The only somewhat arbitrary step is the precise subtraction of the UV parts. The method we chose subtracts some finite parts too, for example, $\log(\mu^2)$ in dimensional regularization, which is a welcome feature, but in principle we could have chosen something else.

Nonetheless, with the counterterm in Eq.~\eqref{eq:ZAfullsimple} we trivially satisfy all the mixing requirements listed in Refs.~\cite{freitas2002,denner2018}, i.e. UV divergences and gauge-dependence are taken care of, there is no process dependence and the scheme is flavour democratic, one can also take the massless and mass degenerate limits at will. With Eq.~\eqref{eq:ZAfullsimple} having all these properties, we consider $\delta Z^A$ to be the main result of this paper.

\subsection{Renormalized hadronic \texorpdfstring{$W$}{W} decay at 1-loop}
\label{sec:Wreno}

Now we turn to the $W^+$ decay into up- and down-type quarks and apply our renormalization scheme of~\cite{drauksas2023} presented in the sections above. The tree level amplitude
\begin{equation}
\begin{split}
    i\mathcal{M}_{0}=&\quad
      \vcenter{\hbox{\begin{tikzpicture}
\begin{feynman}
  \vertex (a) at (-1.5,0);
  \vertex (m) at (0,0);
  \vertex (b) at (1.4,0.9);
  \vertex (c) at (1.4,-0.9);
  \diagram*{
  (a)--[anti charged boson, edge label=$W^-$](m)--[fermion, edge label=$\bar{\psi}_u$](b);
  (c)--[fermion, edge label=$\psi_d$](m);
  };
\end{feynman}
\end{tikzpicture}}}
\\
&\\
=&-i\frac{g V_{ud}}{\sqrt{2}}\bar{u}(p_u)\epsilon_\mu(q) \gamma^\mu P_L v(p_d)\,,
\end{split}
\end{equation}
with $u(p_u)$ and $v(p_d)$ being the external spinors for the up- and anti-down-type quarks, $\epsilon_\mu(q)$ is the polarization vector of the incoming $W$ boson, and $p_{u,d}$, $q$ are their respective momenta. This amplitude leads to the tree-level $W$ decay widths, which are very well known in the literature~\cite{denner1990a,denner1993,kniehl2000}
\begin{equation}
\label{eq:WdecaytreeGamma}
\begin{split}
    \Gamma_0^{ud}=&\frac{\alpha N V^{}_{ud}V^\star_{ud}}{24 s^2 m^5_W}\sqrt{\lambda\left[ m_W^2, m_u^2, m_d^2 \right]}
    \left( 3 m_W^2\left(m_W^2-m_u^2-m_d^2 \right) - \lambda\left[ m_W^2, m_u^2, m_d^2 \right] \right)\,,
\end{split}
\end{equation}
where $N=3$ is the number of colors and $\lambda$ is the K\"{a}llen triangle function~\cite{kallen1964}. At 1-loop the amplitude can be written in terms of 4 form factors, only one of which needs counterterms~\cite{denner1990a,denner1993}
\begin{align}
    i\mathcal{M}_1=& \quad
      \vcenter{\hbox{\begin{tikzpicture}
  \begin{feynman}
    \vertex (a) at (-1.5,0);
    \vertex [blob] (m) at (0,0){1PI};
    \vertex (b) at (1.4,0.9);
    \vertex (c) at (1.4,-0.9);
    \diagram*{
    (a)--[anti charged boson, edge label=$W^-$](m)--[fermion, edge label=$\bar{\psi}_u$](b);
    (c)--[fermion, edge label=$\psi_d$](m);
    };
  \end{feynman}
  \end{tikzpicture}}}
  \hspace{1em}
  +
  \hspace{1em}
  \vcenter{\hbox{\begin{tikzpicture}
  \begin{feynman}
    \vertex (a) at (-1.5,0);
    \vertex [crossed dot] (m) at (0,0){};
    \vertex (b) at (1.4,0.9);
    \vertex (c) at (1.4,-0.9);
    \diagram*{
    (a)--[anti charged boson, edge label=$W^-$](m)--[fermion, edge label=$\bar{\psi}_u$](b);
    (c)--[fermion, edge label=$\psi_d$](m);
    };
  \end{feynman}
  \end{tikzpicture}}}
  \nonumber \\
  & \nonumber \\
  =& -\frac{g V_{ud}}{\sqrt{2}}\left(\sum_{a=1,2}\sum_{\sigma=L,R}iM_a^\sigma\sigma F_a^\sigma+i\delta F_1^L M_1^L \right)\,.
\end{align}
Here
\begin{align}
  M_1^{L,R}&=\bar{u}(p_u) \gamma^\mu P_{L,R} v(p_d)\epsilon_\mu(q)\,,\\
  M_2^{L,R}&=\bar{u}(p_u) P_{L,R} v(p_d)\, p_u^\mu\epsilon_\mu(q)
            \nonumber \\
          &=-\bar{u}(p_u) P_{L,R} v(p_d)\, p_d^\mu\epsilon_\mu(q)\,,
\end{align}
and
\begin{equation}
\begin{split}
      \delta F_1^L =& \frac{\delta g}{g} 
                + \delta Z_W
                + \frac{1}{V_{ud}}\sum_{u^\prime} \delta Z^\dagger_{L,\,uu^\prime} V_{u^\prime d}
                + \frac{1}{V_{ud}}\sum_{d^\prime}V_{ud^\prime} \delta Z_{L,\,d^\prime d} \,,
\end{split}
\end{equation}
with all the counterterms defined in Eqs.~\eqref{eq:gct},~\eqref{eq:AZmassctfinal},~\eqref{eq:diagonalfieldct},~\eqref{eq:offdiagonalfieldct},~\eqref{fermionfieldHerm}, and~\eqref{eq:fermionfieldoffdiag} and the explicit form factors $F_a^\sigma$ can be found in~\cite{denner1990a,denner1993}. We reiterate that there is no counterterm for the CKM mixing matrix in our scheme. The average over $W$ and sum over fermion polarizations leads to the following factors in the squared amplitude
\begin{align}
    G_1^L=&\sum_\text{pols.} \left( M^L_1 \right)^\dagger M^L_1 
    =  3 \left(m_W^2-m_u^2-m_d^2 \right)- \frac{\lambda\left[ m_W^2, m_u^2, m_d^2 \right]}{m_W^2}\,,\\
    G_1^R=&\sum_\text{pols.} \left( M^L_1 \right)^\dagger M^R_1=6m_u m_d\,,\\
    G_2^L=&\sum_\text{pols.} \left( M^L_1 \right)^\dagger M^L_2=-\frac{m_u}{2m_W^2}\lambda\left[ m_W^2, m_u^2, m_d^2 \right]\,,\\
    G_2^R=&\sum_\text{pols.} \left( M^L_1 \right)^\dagger M^R_2=-\frac{m_d}{2m_W^2}\lambda\left[ m_W^2, m_u^2, m_d^2 \right]\,,
\end{align}
with $G_1^L$ already appearing explicitly in $\Gamma_0$ of Eq.~\eqref{eq:WdecaytreeGamma}. Then we may write the one-loop corrected decay width as
\begin{equation}
  \Gamma_1=\Gamma_0\left( 1+\delta_\mathrm{virt}+\delta_\mathrm{real}\right)\,,
\end{equation}
where
\begin{equation}
\begin{split}
  \delta_\mathrm{virt}=& 2\mathrm{Re}\Biggl[
                \frac{1}{G_1^L} \sum_{a=1,2} \sum_{\sigma=L,R}G_a^\sigma F_a^\sigma
                +\frac{\delta g}{g}
                + \delta Z_W+\frac{1}{V_{ud}}\sum_{u^\prime} \delta Z^\dagger_{L,\,uu^\prime} V_{u^\prime d}
                +\frac{1}{V_{ud}}\sum_{d^\prime}V_{ud^\prime} \delta Z_{L,\,d^\prime d}\Biggr] 
\end{split}
\end{equation}
and $\delta_\mathrm{real}$ represents the infrared-divergent corrections coming from bremsstrahlung contributions of the decay $W^+\to u\bar{d}\gamma$. The explicit contributions of $\delta_\mathrm{real}$ may be found in~\cite{denner1990a,denner1993}.

The corrections $\delta_\mathrm{virt,real}$ may be further split into the electroweak and QCD parts
\begin{equation}
  \delta_\mathrm{virt,real}=\delta_\mathrm{virt,real}^\mathrm{EW}+\delta_\mathrm{virt,real}^\mathrm{QCD}\,.
\end{equation}
The electroweak corrections may be reduced by parametrizing the tree-level decay width with the Fermi constant $G_\mathrm{F}$ instead of the fine structure constant as~\cite{sirlin1980,marciano1980,hollik1990,denner1993,pokorski2000,denner2020}
\begin{equation}
  G_\mathrm{F}=\frac{\pi \alpha}{\sqrt{2}s^2 m_W^2}\frac{1}{1-\Delta r}
\end{equation}
with 
\begin{equation}
\begin{split}
\label{eq:Deltar}
  \Delta r =& -\frac{1}{m_W^2}\left( \Pi_{WW}(0)-\Pi_{WW}(m_W^2) \right)
            -\frac{c^2}{s^2}\left( \frac{\Pi_{WW}(m_W^2)}{m_W^2}-\frac{\Pi_{ZZ}(m_Z^2)}{m_Z^2} \right) \\
            &-\left.\frac{\partial \Pi_{AA}(p^2)}{\partial p^2}\right|_{p^2=0}
            +2\frac{c}{s}\frac{\Pi_{ZA}(0)}{m_Z^2}
            +\frac{\alpha}{4\pi s^2}\left( 6 + \left( \frac{7}{2s^2}-2 \right)\ln c^2 \right)
\end{split}
\end{equation}
so that upon reparametrizing Eq.~\eqref{eq:WdecaytreeGamma} we get
\begin{equation}
  \label{eq:WdecaytreeGammaBar}
\begin{split}
    \bar{\Gamma}^{ud}_0\left( 1-\Delta r \right)=&\frac{G_\mathrm{F} N V^{}_{ud}V^\star_{ud}}{12\sqrt{2} \pi m_W}\sqrt{\lambda\left[ m_W^2, m_u^2, m_d^2 \right]} G_1^L\left( 1-\Delta r \right)\,.  
\end{split}
\end{equation}
Note that the last term and part of $\Pi_{ZA}$ in $\Delta r$ come from vertex and box corrections, in addition, $\Delta r$ is UV finite and gauge-independent, but the vertex and box contributions are given in the Feynman gauge, therefore, all the self-energies must be evaluated in the Feynman gauge, too.

Finally, with this parametrization, which is the $G_\mathrm{F}$ input scheme appropriate for charged current processes~\cite{kniehl2000,denner2020}, the partial decay width is~\cite{denner1990a,denner1993}
\begin{equation}
\label{eq:Wdecaywidth1loop}
  \bar{\Gamma}_1^{ud}=\bar{\Gamma}^{ud}_0\left( 1+\delta_\text{virt}+\delta_\text{real}-\Delta r \right)\,,
\end{equation}
where $\delta^\text{EW}_\text{virt}$ and $\Delta r$ nearly cancel out. Note that the Fermi constant appears only in $\bar{\Gamma}_0$, while $\delta_\mathrm{virt,real}$ and $\Delta r$ still contain the fine structure constant $\alpha$.

\section{Numerical results}
\label{sec:numres}

We proceed to the numerical evaluation of Eq.~\eqref{eq:Wdecaywidth1loop} for all the hadronic decay channels of the $W$ boson, with the goal of reproducing and updating Table I found in~\cite{almasy2009err}.

We have recalculated all the virtual loop corrections in $\delta_\mathrm{virt}$ with \texttt{FeynCalc}~\cite{FeynCalc}, \texttt{FeynArts}~\cite{hahn2000}, and \texttt{PackageX}~\cite{patel2015}, with the \texttt{FeynRules}~\cite{alloul2014} SM model files~\cite{duhr2016}. The bremsstrahlung contributions in $\delta_\mathrm{real}$ were also recalculated, where we have found~\cite{naeem2025} and the additional private comments by S.~Groote to be very useful. For the evaluation of virtual corrections we have used \texttt{LoopTools}~\cite{LoopTools} (except for a few $C_0\left[0,a,a,b,b,c\right]$ functions which we evaluated explicitly with \texttt{PackageX} and inserted the numerical values) and made sure to use quadruple precision by writing a few small \texttt{fortran} scripts that call \texttt{LoopTools} and evaluate the relevant Passarino-Veltman functions. The contributions in $\delta_\mathrm{real}$ were evaluated by explicitly inserting the numerical values.

We made sure that all the UV and IR divergences fully cancel both analytically and numerically. We have used the photon mass to regulate the IR divergences, however, \texttt{LoopTools} uses dimensional regularization, so we employed the following relation
\begin{equation}
  \ln m_\gamma^2 \longleftrightarrow \frac{1}{\epsilon^{}_\mathrm{IR}}-\gamma_\mathrm{E}+\ln\left( 4\pi \right)\,,
\end{equation}
which is valid if there are no $1/\epsilon^2_\mathrm{IR}$ divergences. Gauge-independence of our results within the $R_\xi$ gauges was also confirmed analytically\footnote{The decay widths are gauge-dependent in the scheme of~\cite{denner1990}.}. Overall, we have cross-checked our virtual and real corrections analytically and/or numerically with those of~\cite{denner1990a,denner1993} and found no disagreements. We have fully reproduced the numerical results (i.e. all 6 digits) of Tables I and II of~\cite{denner1990} by only using the appropriate numerical values given by the authors\footnote{$m_W$ there is a derived parameter and requires more digits than displayed by the authors.}. On the other hand, the authors of~\cite{almasy2009err} have reproduced the numerical results of~\cite{kniehl2000}, who have cross-checked their result with~\cite{denner1993}. We, too, were able to reproduce the results of the first two columns of Table~9.2 of~\cite{denner1993}, but there only 4 significant digits are displayed. Unfortunately, while we get the same tree-level values as~\cite{kniehl2000}, at 1-loop our results differ from the ones of~\cite{kniehl2000,almasy2009err} by $10^{-7}$ (the 7th digit) in the $W^+\to u\bar{d}$ channel and with the difference increasing to $\sim10^{-4}$ ($\sim$4th digit) in the $W^+\to c\bar{b}$ channel. We did contact the authors of~\cite{kniehl2000,almasy2009err}, but the relevant colleagues have already retired, and the code is no longer available, hence, we were unable to do an in-depth comparison. For reference, we include our results with the input values of~\cite{almasy2009err} in Appendix~\ref{sec:Bcomparing08}.

The numerical inputs we have used are the following~\cite{navas2024}
\begin{equation}
\begin{alignedat}{6}
  \alpha(0)&=1/137.035999178\,,& \quad G_\mathrm{F}&=1.1663788\times 10^{-5}~\mathrm{GeV}^{-2}\,,&\quad \alpha_s(m^{}_Z)&=0.1175\,, \\
  m_W&=80.3692~\mathrm{GeV}\,,& \quad m_Z&=91.1880~\mathrm{GeV}\,, &\quad m_H&=125.20~\mathrm{GeV}\,,\\
  m_e&=0.51099895000~\mathrm{MeV}\,,&\quad m_\mu&=0.1056583755~\mathrm{GeV}\,, & \quad m_\tau& = 1.77693~\mathrm{GeV}\,,\\
  m_u&=2.16\times10^{-3}~\mathrm{GeV}\,, &\quad m_c&=1.2730~\mathrm{GeV}\,,& \quad m_t&=172.57~\mathrm{GeV}\,, \\
  m_d&=4.70\times 10^{-3}~\mathrm{GeV}\,,&\quad m_s&=93.5\times 10^{-3}~\mathrm{GeV}\,,&\quad m_b&=4.183~\mathrm{GeV}\,.
\end{alignedat}
\end{equation}

For the $W$ decay it is appropriate to use the strong coupling constant $\alpha_s$ at $m^{}_W$ instead of $m^{}_Z$~\cite{kniehl2000,kara2013}
\begin{equation}
  \alpha_s(m^{}_W)=\frac{\alpha_s(m^{}_Z)}{1+\frac{\alpha_s(m^{}_Z)\beta_0}{\pi} \ln(m_W^2/m_Z^2)}= 0.1197
\end{equation}
with $\beta_0=11/4-n_f/6$ and with the number of flavors $n_f=5$ (we display only 4 digits, but employ all available digits in the evaluation).
For the CKM matrix we use the Wolfenstein parametrization~\cite{wolfenstein1983,buras1994} with
\begin{equation}
  \lambda=0.22501\,, \quad A=0.826\,, \quad \bar{\rho}=0.1591\,,\quad \bar{\eta}=0.3523\,.
\end{equation}

The resulting values of the CKM elements to 3 significant digits are
\begin{equation}
\begin{alignedat}{6}
  V_{ud}&=0.974\,, &\quad V_{us} &=0.225\,, &\quad V_{ub} & = (1.53 - i 3.40)\times10^{-3}\,, \\
  V_{cd}&=-0.225-i1.39\times10^{-4}\,,&\quad V_{cs}&= 0.973-i3.20\times10^{-5}\,,&\quad V_{cb}&=0.0418\,, \\
  V_{td}&=\left(7.92-i3.31 \right)\times10^{-3}\,,&\quad V_{ts}&=-0.0411-i7.65\times10^{-4}\,,&\quad V_{tb}&=0.999\,.
\end{alignedat}
\end{equation}

With these inputs we evaluate Eq.~\eqref{eq:Deltar} and get
\begin{equation}
  \label{eq:Deltarnum}
  \Delta r = 0.03883828\,. 
\end{equation}
Further, for the tree-level decay widths we evaluate $\bar{\Gamma}_0^{ud}$ of Eq.~\eqref{eq:WdecaytreeGammaBar}
\begin{align}
    \bar{\Gamma}_0^{ud}&= 6.4690927\times 10^{-1}~\mathrm{GeV}\,, 
    \nonumber \\
    \bar{\Gamma}_0^{us}&= 3.4499309\times 10^{-2}~\mathrm{GeV}\,, 
    \nonumber \\
    \bar{\Gamma}_0^{ub}&=9.4470082\times10^{-6}~\mathrm{GeV}\,,
    \nonumber \\
    \bar{\Gamma}_0^{cd}&=3.4445715\times 10^{-2}~\mathrm{GeV}\,, 
    \nonumber \\
    \bar{\Gamma}_0^{cs}&=6.4552340\times 10^{-1}~\mathrm{GeV}\,,
    \nonumber \\
    \bar{\Gamma}_0^{cb}&=1.1864313\times 10^{-3}~\mathrm{GeV}\,,   
    \end{align}
which add up to the total hadronic decay width
\begin{equation}
  \bar{\Gamma}_0^\mathrm{had}=\sum_{u^\prime,d^\prime}\bar{\Gamma}_0^{u^\prime d^\prime}=1.3625736~\mathrm{GeV}\,.
\end{equation}

For 1-loop results we provide Tables~\ref{tab:normalschemes24} and~\ref{tab:otherschemes24}, where the numerical results are given for CKM matrix renormalization schemes found in~\cite{denner1990,gambino1999,diener2001,kniehl2006,kniehl2009a}, the $\overline{\mathrm{MS}}$ scheme (only CKM is renormalized $\overline{\mathrm{MS}}$, everything else is on-shell), the scheme where there is no mixing on external legs and the CKM counterterm is trivial, and the first results for the scheme in~\cite{drauksas2023}, which we also outlined in this work. Note, that the results of~\cite{denner1990} are gauge-dependent and were computed in the Feynman gauge. It is evident that the effects of the CKM matrix renormalization are rather small, therefore, just like in~\cite{kniehl2000,almasy2009err}, we display more digits than the actual accuracy achieved by the calculation (i.e. one would need to include higher order corrections). The results in all the schemes numerically (up to displayed digits) are nearly identical. On the other hand, the differences might become important if the scheme is applied to neutrinos or other beyond the Standard Model physics. At the very least, the scheme we have presented in this work is numerically sound and stands on an equal footing with the other schemes in the literature. 

\begin{table}[h!]
\centering
\renewcommand{\arraystretch}{1.3}
\begin{tabular}{|l|c|c|c|c|c|c|}
\hline\hline
\multirow{2}{1.6cm}{\textbf{Partial widths}}
& \multirow{2}{1.9cm}{\textbf{Ref.~\cite{denner1990}}} 
& \multirow{2}{1.9cm}{\textbf{Ref.~\cite{gambino1999}}} 
& \multirow{2}{1.9cm}{\textbf{Ref.~\cite{dienerkniehl}} }
& \multirow{2}{1.9cm}{\textbf{Ref.~\cite{kniehl2006}} }
& \multirow{2}{1.9cm}{\textbf{Ref.~\cite{kniehl2009a}}}
& \multirow{2}{1.9cm}{\centering \textbf{This work~\cite{drauksas2023}}}  
\\
&  &  &  &  &  & \\
\hline

$\bar{\Gamma}^{ud}\times 10$
& 6.6919100 & 6.6919100 & 6.6919100 & 6.6919100 & 6.6919100 & 6.6919100\\

$\bar{\Gamma}^{us}\times 10^2$
& 3.5687660 & 3.5687659 & 3.5687659 & 3.5687659 & 3.5687659 & 3.5687660\\

$\bar{\Gamma}^{ub}\times 10^6$
& 9.7927690 & 9.7928151 & 9.7928647 & 9.7928151 & 9.7928151 & 9.7927572\\

$\bar{\Gamma}^{cd}\times 10^2$
& 3.5641478 & 3.5641478 & 3.5641478 & 3.5641478 & 3.5641478 & 3.5641478\\

$\bar{\Gamma}^{cs}\times 10$
& 6.6793383 & 6.6793382 & 6.6793382 & 6.6793382 & 6.6793382 & 6.6793383\\

$\bar{\Gamma}^{cb}\times 10^3$
& 1.2301785 & 1.2301843 & 1.2301906 & 1.2301843 & 1.2301843 & 1.2301771\\
\hline
$\bar{\Gamma}^\mathrm{had}$
& 1.4096939 & 1.4096939 & 1.4096939 & 1.4096939 & 1.4096939 & 1.4096939\\
\hline\hline
\end{tabular}
\caption{Partial and total hadronic $W$ decay widths (in $\mathrm{GeV}$) in various CKM renormalization schemes found in the literature.}
\label{tab:normalschemes24}
\end{table}

\begin{table}[h!]
\centering
\renewcommand{\arraystretch}{1.3}
\begin{tabular}{|l|c|c|}
\hline\hline
\multirow{2}{1.6cm}{\textbf{Partial widths}}
& \multirow{2}{2cm}{\centering $\overline{\text{MS}}$ \textbf{scheme}}
& \multirow{2}{2cm}{$ V^\mathrm{ext.}_{ji}=\delta_{ji}$ }
\\
&  &  \\
\hline

$\bar{\Gamma}^{ud}\times 10$
& 6.6919067 & 6.6919100 \\

$\bar{\Gamma}^{us}\times 10^2$
& 3.5687925 & 3.5687659\\

$\bar{\Gamma}^{ub}\times 10^6$
& 9.8653150 & 9.7928151 \\

$\bar{\Gamma}^{cd}\times 10^2$
& 3.5641431 & 3.5641478\\

$\bar{\Gamma}^{cs}\times 10$
& 6.6792473 & 6.6793382\\

$\bar{\Gamma}^{cb}\times 10^3$
& 1.2392896 & 1.2301843\\
\hline
$\bar{\Gamma}^\mathrm{had}$
& 1.4096939 & 1.4096939 \\
\hline\hline
\end{tabular}
\caption{Partial and total hadronic $W$ decay widths (in $\mathrm{GeV}$) in the $\overline{\mathrm{MS}}$ renormalization scheme for the CKM matrix and a scheme where there is no mixing on external legs (trivial CKM counterterm).}
\label{tab:otherschemes24}
\end{table}

For a more convenient comparison we also provide the virtual and real correction split into EW and QCD parts for the initial scheme of~\cite{denner1990} and ours~\cite{drauksas2023} in Table~\ref{tab:deltavirtb24}. The differences between the two schemes are rather small numerically and are only in the electroweak part as expected. The difference is effectively caused by the gauge-dependent contributions in the Feynman gauge, which are present in~\cite{denner1990}, but are absent in our scheme~\cite{drauksas2023}. Here it is also evident that $\Delta r$ in Eq.~\eqref{eq:Deltarnum} nearly completely cancels the electroweak corrections in Eq.~\eqref{eq:Wdecaywidth1loop} so that the resulting EW corrections are roughly $-0.3\%$ as should be the case. The QCD corrections range between $3.8$ and $4.05$ percent.

\begin{table*}[h!]
\centering
\renewcommand{\arraystretch}{1.3}
\begin{tabular}{|l|c|c|c|c|}
\hline\hline
\multirow{2}{*}{\textbf{Channel}}
& \multicolumn{2}{c|}{\centering \textbf{Ref.~\cite{denner1990}}} 
& \multicolumn{2}{c|}{\centering \textbf{This work~\cite{drauksas2023}}}  
\\
\cline{2-5}
& $\delta^\mathrm{EW}_\mathrm{virt}+\delta^\mathrm{EW}_\mathrm{real}$ & $\delta^\mathrm{QCD}_\mathrm{virt}+\delta^\mathrm{QCD}_\mathrm{real}$ & $\delta^\mathrm{EW}_\mathrm{virt}+\delta^\mathrm{EW}_\mathrm{real}$ & $\delta^\mathrm{QCD}_\mathrm{virt}+\delta^\mathrm{QCD}_\mathrm{real}$ \\
\hline

$W^+\to u\bar{d}$
& 3.5190522 & 3.8091129 & 3.5190522 & 3.8091129 \\

$W^+\to u\bar{s}$
& 3.5190536 & 3.8093366 & 3.5190537 & 3.8093366 \\

$W^+\to u\bar{b}$
& 3.5199894 & 4.0238412 & 3.5198645 & 4.0238412 \\

$W^+\to c\bar{d}$
& 3.5195129 & 3.8357576 & 3.5195131 & 3.8357576 \\

$W^+\to c\bar{s}$
& 3.5195143 & 3.8359814 & 3.5195146 & 3.8359814 \\

$W^+\to c\bar{b}$
& 3.5204515 & 4.0506764 & 3.5203264 & 4.0506764 \\

\hline\hline
\end{tabular}
\caption{Electroweak and QCD corrections (in \%) to the hadronic $W$ decay channels in the renormalization schemes of~\cite{denner1990} and~\cite{drauksas2023}. Here $\Delta r$ is not subtracted from the EW corrections.}
\label{tab:deltavirtb24}
\end{table*}

\pagebreak

\section{Discussion \& Conclusions}
\label{sec:conclusions}

In this work we have considered the problem of the renormalization of particle mixing. We chose an approach where instead of mixing matrix counterterms one introduces the non-diagonal mass counterterms, which then renormalize the mixing. Such an approach arises rather naturally if one considers the properties of basis transformations. Mixing matrix counterterms single out a specific basis, which should have no physical significance whatsoever, hence, these counterterms are inconsistent in this sense. We have managed to implement this approach in the gauge sector of the Standard Model rather easily and instead of renormalizing the Weinberg angle, we have introduced the off-diagonal counterterm in the $Z$ boson and photon mass matrix. The constraints of the gauge sector then allowed to express this mass counterterm in terms of the $W$ and $Z$ boson mass counterterms in the On-Shell scheme. The fermion sector proved to be more difficult.

In the fermion sector, the On-Shell conditions provide a relation between mass and the anti-hermitian part of field renormalization counterterms in terms of the self-energy. At first sight, the relation is degenerate so that one cannot solve for both of the counterterms. Already in Ref.~\cite{drauksas2023} we have suggested employing the mass structures to define the mass and field counterterms, however, a satisfactory practical method was missing until now. In Section~\ref{sec:method} we have used a more fine-grained decomposition of the self-energy at 1-loop, separated the on-shell scalar functions into parts respecting or violating the pseudo-hermiticity constraints, and subtracted the UV divergences from two relevant scalar functions of the extended decomposition. That was enough to find the relevant mass structure and to develop a useful prescription for the anti-hermitian part of the field renormalization in Eq.~\eqref{eq:ZAfullsimple}. With this prescription the particle mixing is then renormalized solely in terms of the 2-point functions in a model- and process-independent way and satisfies all the mixing requirements known to the author~\cite{freitas2002,denner2018}. In this paper we only provided the 1-loop prescription, but we discuss the possibility of extension to higher orders in Appendix~\ref{secA:extensions}, which is, however, beyond the scope of this paper.

For a numerical example we have provided a table with partial and total hadronic $W$ decay widths in the Standard Model in the spirit of~\cite{almasy2009err} with the extension of the very first numerical results for the quark mixing matrix renormalization scheme found in~\cite{drauksas2023} and refined in this work. We have found, that numerically the scheme produces very similar results as compared to other schemes found in the literature. While the example is not phenomenologically relevant, it does show that the scheme does work just as well numerically.

\section*{Acknowledgements}
The author thanks S.~Groote for helpful comments and additional notes regarding~\cite{naeem2025} and IR divergent integrals. The author thanks local colleagues T.~Gajdosik, V.~Dūdėnas, and U.~Igaris for reading the manuscript and suggesting various improvements.\\
This research has been carried out in the framework of the agreement of Vilnius University with the Lithuanian Research Council No.~VS-13.



\appendix

\section{Nielsen Identities and fermion self-energies}
\subsection{Nielsen identities for the fine-grained decomposition}
\label{sec:AnielsenSE}
In this appendix we see how the Nielsen identities~\cite{nielsen1975,gambino2000} are related to the fine-grained decomposition of the fermion self-energies in Eq.~\eqref{eq:selfextradeco}. The Nielsen identities allow taking gauge-parameter derivatives (in $R_\xi$ gauges) of the self-energies:
\begin{equation}
  \label{eq:nielsenfirst}
  \partial_\xi \Sigma_{ji}(\cancel{p})=\sum_k\Lambda_{jk}(\cancel{p})\left[(\cancel{p}-m_i)\delta_{ki}+\Sigma_{ki}(\cancel{p})\right]
  +\sum_k\left[\Sigma_{jk}(\cancel{p})+(\cancel{p}-m_j)\delta_{jk}\right]\bar{\Lambda}_{ki}(\cancel{p})\,,
\end{equation}
where $\Lambda_{ji}=-\Gamma_{\chi\bar{\psi}_j\eta_{\psi_i}}$, $\bar{\Lambda}_{ji}=-\Gamma_{\chi\bar{\eta}_{\psi_j}\psi_i}$ with $\chi$ the BRST source for the gauge parameter $\xi$, $\eta$ and $\bar{\eta}$ the sources for the BRST transformations of the corresponding fermion fields. Here and up to Eq.~\eqref{eq:Nielsen4} we briefly consider $\Sigma$ to contain contributions from all orders.

Since $\Lambda$s have Dirac structure and can be decomposed just as the self-energies in Eq.~\eqref{eq:selfdeco}, we can find the following relations
\begin{align}
  \label{eq:Nielsen1}
  \partial_\xi \Sigma_{ji}^{\gamma L}(p^2)=&
          \Lambda^{sR}_{ji}
          -m_i\Lambda^{\gamma L}_{ji} 
          +\bar{\Lambda}^{sL}_{ji}
          -m_j \bar{\Lambda}^{\gamma L}_{ji}
  \nonumber \\
          &+\Lambda^{sR}_{jk}\Sigma_{ki}^{\gamma L}
          +\Lambda^{\gamma L}_{jk}\Sigma_{ki}^{sL} 
          +\Sigma_{jk}^{\gamma L} \bar{\Lambda}^{sL}_{ki}
          +\Sigma_{jk}^{sR} \bar{\Lambda}^{\gamma L}_{ki}\,,
  \\
  \partial_\xi \Sigma_{ji}^{\gamma R}(p^2)=&
          \Lambda^{sL}_{ji}
          -m_i\Lambda^{\gamma R}_{ji} 
          +\bar{\Lambda}^{sR}_{ji}
          -m_j\bar{\Lambda}^{\gamma R}_{ji}
  \nonumber\\
          &+\Lambda^{sL}_{jk}\Sigma_{ki}^{\gamma R}
          +\Lambda^{\gamma R}_{jk}\Sigma_{ki}^{sR} 
          +\Sigma_{jk}^{\gamma R} \bar{\Lambda}^{sR}_{ki}
          +\Sigma_{jk}^{sL} \bar{\Lambda}^{\gamma R}_{ki}\,,
  \\
  \partial_\xi \Sigma_{ji}^{sL}(p^2)=&
        p^2\Lambda^{\gamma R}_{ji} 
        -m_i\Lambda^{sL}_{ji} 
        +p^2\bar{\Lambda}^{\gamma L}_{ji}
        -m_j\bar{\Lambda}^{sL}_{ji}
  \nonumber\\
        &+p^2\Lambda^{\gamma R}_{jk}\Sigma_{ki}^{\gamma L} 
        +\Lambda^{sL}_{jk}\Sigma_{ki}^{sL} 
        +p^2\Sigma_{jk}^{\gamma R} \bar{\Lambda}^{\gamma L}_{ki}
        +\Sigma_{jk}^{sL} \bar{\Lambda}^{sL}_{ki}\,,
  \\
  \partial_\xi \Sigma_{ji}^{sR}(p^2)=&
        p^2\Lambda^{\gamma L}_{ji} 
        -m_i\Lambda^{sR}_{ji} 
        +p^2\bar{\Lambda}^{\gamma R}_{ji}
        -m_j\bar{\Lambda}^{sR}_{ji}
  \nonumber\\
        &+p^2\Lambda^{\gamma L}_{jk}\Sigma_{ki}^{\gamma R} 
        +\Lambda^{sR}_{jk}\Sigma_{ki}^{sR} 
        +p^2\Sigma_{jk}^{\gamma L} \bar{\Lambda}^{\gamma R}_{ki}
        +\Sigma_{jk}^{sR} \bar{\Lambda}^{sR}_{ki}\,,
        \label{eq:Nielsen4}
\end{align}
where we have dropped the momentum dependence and the sum symbol on the r.h.s. 

We switch to 1-loop and move to the fine-grained decomposition, which is available both for the self-energies and the $\Lambda$s, but, crucially, the $\times$-free $\Lambda$s are at most linear in the external masses. This can be expected since only one of the external legs in the correlation function is that of a physical field, the other two being the BRST sources. Therefore, we have
\begin{equation}
  \label{eq:nielsenextra}
  \begin{split}
  \Lambda^{X}_{ji}=&\Lambda^{X,\times\times}_{ji}
                    +\Lambda^{X,\circ\circ}_{ji}
                    +m_j\Lambda^{X,\bullet\circ}_{ji}\,,
  \\
  \bar{\Lambda}^{X}_{ji}=&\bar{\Lambda}^{X,\times\times}_{ji}
                    +\bar{\Lambda}^{X,\circ\circ}_{ji}
                    +m_i\bar{\Lambda}^{X,\circ\bullet}_{ji}\,.
  \end{split}
\end{equation}
The $\Lambda$s have three external legs and in principle should have three bullets indicating the legs, so here the two bullets rather indicate to which index the mass in the decomposition corresponds. 

We may employ the decompositions of Eq.~\eqref{eq:nielsenextra} and Eq.~\eqref{eq:selfextradeco} to find the gauge derivatives of the scalar self-energy functions. While there are quite a few functions, it is straightforward to decompose the functions by considering the masses. Therefore, one finds the following contributions to the gauge-derivatives of the scalar self-energy functions at 1-loop
\begin{align}
  \label{eq:treesigmalambda1}
  \gamma L:\qquad \partial_\xi \Sigma^{\gamma L,\times\times}_{ji}(p^2)=&
              \Lambda^{sR,\times\times}_{ji}
              -m_i \Lambda^{\gamma L,\times\times}_{ji}
              +\bar{\Lambda}^{sL,\times\times}_{ji}
              -m_j\bar{\Lambda}^{\gamma L,\times\times}_{ji}\,,
  \\
  \partial_\xi \Sigma^{\gamma L,\circ\circ}_{ji}(p^2)=&
              \Lambda^{sR,\circ\circ}_{ji}
              +\bar{\Lambda}^{sL,\circ\circ}_{ji}\,,
  \\
  \partial_\xi \Sigma^{\gamma L,\circ\bullet}_{ji}(p^2)=&
              -\Lambda^{\gamma L,\circ\circ}_{ji}
              +\bar{\Lambda}^{sL,\circ\bullet}_{ji} \,, 
  \\
  \partial_\xi \Sigma^{\gamma L,\bullet\circ}_{ji}(p^2)=&
               \Lambda^{sR,\bullet\circ}_{ji}
              -\bar{\Lambda}^{\gamma L,\circ\circ}_{ji}\,,
  \\
  \partial_\xi \Sigma^{\gamma L,\bullet\bullet}_{ji}(p^2)=&
              -\Lambda^{\gamma L,\bullet\circ}_{ji}
              -\bar{\Lambda}^{\gamma L,\circ\bullet}_{ji}\,,
  \\
 \gamma R:\qquad   \partial_\xi \Sigma^{\gamma R,\times\times}_{ji}(p^2)=&
              \Lambda^{sL,\times\times}_{ji}
              -m_i \Lambda^{\gamma R,\times\times}_{ji}
              +\bar{\Lambda}^{sR,\times\times}_{ji}
              -m_j\bar{\Lambda}^{\gamma R,\times\times}_{ji}\,,
  \\
  \partial_\xi \Sigma^{\gamma R,\circ\circ}_{ji}(p^2)=&
              \Lambda^{sL,\circ\circ}_{ji}
              +\bar{\Lambda}^{sR,\circ\circ}_{ji}\,,
  \\
  \partial_\xi \Sigma^{\gamma R,\circ\bullet}_{ji}(p^2)=&
              -\Lambda^{\gamma R,\circ\circ}_{ji}
              +\bar{\Lambda}^{sR,\circ\bullet}_{ji}\,,
  \\
  \partial_\xi \Sigma^{\gamma R,\bullet\circ}_{ji}(p^2)=&
               \Lambda^{sL,\bullet\circ}_{ji}
              -\bar{\Lambda}^{\gamma R,\circ\circ}_{ji}\,,
  \\
  \partial_\xi \Sigma^{\gamma R,\bullet\bullet}_{ji}(p^2)=&
              -\Lambda^{\gamma R,\bullet\circ}_{ji}
              -\bar{\Lambda}^{\gamma R,\circ\bullet}_{ji}\,,
  \\
 sL:\qquad   \partial_\xi \Sigma^{sL,\times\times}_{ji}(p^2)=&
              p^2\Lambda^{\gamma R,\times\times}_{ji}
              -m_i \Lambda^{sL,\times\times}_{ji}
              +p^2\bar{\Lambda}^{\gamma L,\times\times}_{ji}
              -m_j\bar{\Lambda}^{sL,\times\times}_{ji}\,,
  \\
  \partial_\xi \Sigma^{sL,\circ\circ}_{ji}(p^2)=&
              p^2 \Lambda^{\gamma R,\circ\circ}_{ji}  
              +p^2\bar{\Lambda}^{\gamma L,\circ\circ}_{ji}\,,
  \\
  \partial_\xi \Sigma^{sL,\circ\bullet}_{ji}(p^2)=&
              -\Lambda^{sL,\circ\circ}_{ji}
              +p^2\bar{\Lambda}^{\gamma L,\circ\bullet}_{ji}\,,
  \\
  \partial_\xi \Sigma^{sL,\bullet\circ}_{ji}(p^2)=&
              p^2 \Lambda^{\gamma R,\bullet\circ}_{ji}
              -\bar{\Lambda}^{sL,\circ\circ}_{ji}\,,
  \\
  \partial_\xi \Sigma^{sL,\bullet\bullet}_{ji}(p^2)=&
              -\Lambda^{sL,\bullet\circ}_{ji}
              -\bar{\Lambda}^{sL,\circ\bullet}_{ji}\,,
  \\
  sR:\qquad  \partial_\xi \Sigma^{sR,\times\times}_{ji}(p^2)=&
              p^2\Lambda^{\gamma L,\times\times}_{ji}
              -m_i \Lambda^{sR,\times\times}_{ji}
              +p^2\bar{\Lambda}^{\gamma R,\times\times}_{ji}
              -m_j\bar{\Lambda}^{sR,\times\times}_{ji}\,,
  \\
  \partial_\xi \Sigma^{sR,\circ\circ}_{ji}(p^2)=&
              p^2 \Lambda^{\gamma L,\circ\circ}_{ji}  
              +p^2\bar{\Lambda}^{\gamma R,\circ\circ}_{ji}\,,
  \\
  \partial_\xi \Sigma^{sR,\circ\bullet}_{ji}(p^2)=&
              -\Lambda^{sR,\circ\circ}_{ji}
              +p^2\bar{\Lambda}^{\gamma R,\circ\bullet}_{ji}\,,
  \\
  \partial_\xi \Sigma^{sR,\bullet\circ}_{ji}(p^2)=&
              p^2 \Lambda^{\gamma L,\bullet\circ}_{ji}
              -\bar{\Lambda}^{sR,\circ\circ}_{ji}\,,
  \\
  \partial_\xi \Sigma^{sR,\bullet\bullet}_{ji}(p^2)=&
              -\Lambda^{sR,\bullet\circ}_{ji}
              -\bar{\Lambda}^{sR,\circ\bullet}_{ji}\,.
  \label{eq:treesigmalambda36}
\end{align}

\subsection{Hints on renormalization beyond 1-loop}
\label{secA:extensions}

In section~\ref{sec:method} in Eqs.~\eqref{eq:firstlinedmla}--\eqref{eq:secondlinedmlb} we give gauge-independent combinations of the self-energy scalar functions at 1-loop and, when taken on-shell, we consider these to be the contributions to the off-diagonal mass counterterms. These combinations are not difficult to find at 1-loop since the Nielsen identity is not too complicated, but going beyond 1-loop may seem cumbersome. However, there are a couple of hints that suggest otherwise. 

The hints come from seeing how the renormalization of lower orders impacts the higher ones, but for that we need to adopt some notation of~\cite{drauksas2023}. We denote the self-energy computed with bare parameters as $\widetilde{\Sigma}(\cancel{p})$, for example, at 1st order $\widetilde{\Sigma}^{(1)}(\cancel{p})$ consists of 1-loop 1-particle-irreducible (1PI) diagrams and the 1-loop mass counterterm, at the 2nd order $\widetilde{\Sigma}^{(2)}(\cancel{p})$ consists of 2-loop 1PI diagrams, 1-loop 1PI diagrams that have 1-loop counterterms and the 2-loop mass counterterm. The renormalized self-energy is simply given by 
\begin{equation}
\Sigma^R=\gamma^0Z^\dagger\gamma^0 \widetilde{\Sigma}Z\,.
\end{equation}
As a cross-check, the renormalized self-energy at 1st order gives Eq.~\eqref{eq:renormalizedSigma}, but here we do not expand all the parameters.

Having the renormalized self-energy we may impose the On-Shell renormalization conditions at 1- and 2-loops. At 1-loop the no-mixing condition gives
\begin{equation}
\Sigma^R_{ji}u_i=0 
\quad \Rightarrow \quad
\delta Z^{(1)}_{ji}u_i=-\frac{1}{\cancel{p}-m_j}\widetilde{\Sigma}^{(1)}_{ji}u_i
\quad \text{for } i\neq j \,,
\end{equation}
while from the unit residue condition we have
\begin{equation}
\lim_{\cancel{p}\to m_i}\frac{1}{\cancel{p}-m_i}\Sigma^R_{ii}u_i=u_i
\quad \Rightarrow \quad
\delta Z^{H,(1)}_{(ii)}u_i=-\lim_{\cancel{p}\to m_i}\frac{1}{\cancel{p}-m_i}\widetilde{\Sigma}^{(1)}_{ii}u_i\,,
\end{equation}
where $\delta Z^H$ is the hermitian part of the field renormalization. Effectively, both conditions provide the same form for the 1st order field renormalization and we may write
\begin{equation}
  \label{eq:firstorderdZ}
\delta Z^{(1)}_{ji}u_i=-\frac{1}{\cancel{p}-m_j}\widetilde{\Sigma}^{(1)}_{ji}u_i
\end{equation}
for all $i$ and $j$.

At 2nd order the renormalization conditions provide the following
\begin{equation}
  \begin{split}
    \delta Z^{(2)}_{ji}u_i=&-\frac{1}{\cancel{p}-m_j}\left(
      \widetilde{\Sigma}^{(2)}_{ji}
      +\widetilde{\Sigma}^{(1)}_{jk}\delta Z^{(1)}_{ki}
    \right)u_i\\
    =&-\frac{1}{\cancel{p}-m_j}\left(
      \widetilde{\Sigma}^{(2)}_{ji}
      -\widetilde{\Sigma}^{(1)}_{jk}\frac{1}{\cancel{p}-m_k}\widetilde{\Sigma}^{(1)}_{ki}
    \right)u_i\,.
  \end{split}
\end{equation}
As mentioned, $\widetilde{\Sigma}^{(2)}$ contains the second order mass counterterm $\delta m^{(2)}$ so that this is the equivalent of Eq.~\eqref{eq:fermionfieldmassrelation} albeit in a slightly different form. We may consider the gauge-dependence of the 2nd order field renormalization by using Eq.~\eqref{eq:nielsenfirst} for the $\widetilde{\Sigma}$ self-energies and get
\begin{equation}
  \begin{split}
    \partial_\xi\delta Z^{(2)}_{ji}u_i=&
    -\frac{1}{\cancel{p}-m_j}\Bigg(
      \partial_\xi\widetilde{\Sigma}^{(2)}_{ji}
      -\partial_\xi\widetilde{\Sigma}^{(1)}_{jk}\frac{1}{\cancel{p}-m_k}\widetilde{\Sigma}^{(1)}_{ki}
      -\widetilde{\Sigma}^{(1)}_{jk}\frac{1}{\cancel{p}-m_k}\partial_\xi\widetilde{\Sigma}^{(1)}_{ki}
    \Bigg)u_i\\
    =&-\frac{1}{\cancel{p}-m_j}\Bigg(
      \widetilde{\Lambda}^{(2)}_{ji}(\cancel{p}-m_i)
  +\widetilde{\Lambda}^{(1)}_{jk}\widetilde{\Sigma}^{(1)}_{ki}
  +(\cancel{p}-m_j)\widetilde{\bar{\Lambda}}^{(2)}_{ji}
  +\widetilde{\Sigma}^{(1)}_{jk}\widetilde{\bar{\Lambda}}^{(1)}_{ki}\\
      &-\widetilde{\Lambda}^{(1)}_{jk}\widetilde{\Sigma}^{(1)}_{ki}
      -(\cancel{p}-m_j)\widetilde{\bar{\Lambda}}^{(1)}_{jk}\frac{1}{\cancel{p}-m_k}\widetilde{\Sigma}^{(1)}_{ki}
      -\widetilde{\Sigma}^{(1)}_{jk}\frac{1}{\cancel{p}-m_k}\widetilde{\Lambda}^{(1)}_{ki}(\cancel{p}-m_i)
      -\widetilde{\Sigma}^{(1)}_{jk}\widetilde{\bar{\Lambda}}^{(1)}_{ki}
    \Bigg)u_i\\
    =&-\frac{1}{\cancel{p}-m_j}\Bigg(
       \widetilde{\Lambda}^{(2)}_{ji}(\cancel{p}-m_i)
       -\widetilde{\Sigma}^{(1)}_{jk}\frac{1}{\cancel{p}-m_k}\widetilde{\Lambda}^{(1)}_{ki}(\cancel{p}-m_i)\\
      &+(\cancel{p}-m_j)\widetilde{\bar{\Lambda}}^{(2)}_{ji}
      -(\cancel{p}-m_j)\widetilde{\bar{\Lambda}}^{(1)}_{jk}\frac{1}{\cancel{p}-m_k}\widetilde{\Sigma}^{(1)}_{ki}
      \Bigg)u_i\\
    =&-\left(\widetilde{\bar{\Lambda}}^{(2)}_{ji}
      +\widetilde{\bar{\Lambda}}^{(1)}_{jk}\delta Z_{ki}^{(1)}
      \right)u_i\,.
  \end{split}
\end{equation}
To get to the last equality we have used the Dirac equation and Eq.~\eqref{eq:firstorderdZ}. The last equality is the second order of the general result present in Eq.~(149) of~\cite{drauksas2023}, however, the more interesting result is before the last equality. It implies that the combination of self-energies
\begin{equation}
  \mathfrak{S}^{(2)}_{ji}=\widetilde{\Sigma}^{(2)}_{ji}
      -\widetilde{\Sigma}^{(1)}_{jk}\frac{1}{\cancel{p}-m_k}\widetilde{\Sigma}^{(1)}_{ki}
\end{equation}
obeys the Nielsen identity, which can be written as
\begin{equation}
  \partial_\xi \mathfrak{S}^{(2)}_{ji} = \mathfrak{L}^{(2)}_{ji}(\cancel{p}-m_i)
                                          +(\cancel{p}-m_j)\bar{\mathfrak{L}}^{(2)}_{ji}\,,
\end{equation}
where we have defined
\begin{equation}
  \begin{split}
    \mathfrak{L}^{(2)}_{ji}=&\widetilde{\Lambda}^{(2)}_{ji}
                            -\widetilde{\Sigma}^{(1)}_{jk}
                            \frac{1}{\cancel{p}-m_k}
                            \widetilde{\Lambda}^{(1)}_{ki}\,,\\
    \bar{\mathfrak{L}}^{(2)}_{ji}=&\widetilde{\bar{\Lambda}}^{(2)}_{ji}
                                  -\widetilde{\bar{\Lambda}}^{(1)}_{jk}
                                  \frac{1}{\cancel{p}-m_k}
                                  \widetilde{\Sigma}^{(1)}_{ki}\,.
  \end{split}
\end{equation}

By iterating this procedure one notices that the field renormalization produces the reducible terms of the Dyson series at every order such that the Nielsen identity keeps its simple form. Having this in mind we give a more concrete definition of $\mathfrak{S}$ as
\begin{equation}
  \mathfrak{S}\equiv i(\cancel{p}-m)
                  \frac{1}{\cancel{p}-m+\widetilde{\Sigma}}
                  i(\cancel{p}-m)\,,
\end{equation}
which is just the amputated 2-point Green's function. Taking the gauge derivative of $\mathfrak{S}$ and using Eq.~\eqref{eq:nielsenfirst} we get
\begin{equation}
  \begin{split}
  \partial_\xi \mathfrak{S}=&(\cancel{p}-m)
                              \frac{1}{\cancel{p}-m+\widetilde{\Sigma}}
                                \left(\widetilde{\Lambda}\left(\cancel{p}-m+\widetilde{\Sigma}\right)
                                +\left( \cancel{p}-m+\widetilde{\Sigma} \right)
                                \widetilde{\bar{\Lambda}}\right)
                              \frac{1}{\cancel{p}-m+\widetilde{\Sigma}}
                              (\cancel{p}-m)\\
                          =&\frac{1}{1+\widetilde{\Sigma}\frac{1}{\cancel{p}-m}}\widetilde{\Lambda}(\cancel{p}-m)
                          +(\cancel{p}-m)\widetilde{\bar{\Lambda}}\frac{1}{1+\frac{1}{\cancel{p}-m}\widetilde{\Sigma}}\,,
  \end{split}
\end{equation}
which can be written in a simple form
\begin{equation}
  \label{eq:Nielsensimple}
  \partial_\xi\mathfrak{S}_{ji}=\mathfrak{L}_{ji}(\cancel{p}-m_i)
                                +(\cancel{p}-m_j)\bar{\mathfrak{L}}_{ji}\,,
\end{equation}
where 
\begin{equation}
    \mathfrak{L}=\frac{1}{1+\widetilde{\Sigma}\frac{1}{\cancel{p}-m}}\widetilde{\Lambda}
    \qquad \text{and} \qquad
    \bar{\mathfrak{L}}=\widetilde{\bar{\Lambda}}\frac{1}{1+\frac{1}{\cancel{p}-m}\widetilde{\Sigma}}\,.
\end{equation}

The upshot is that both the renormalization conditions and the Nielsen identity of $\mathfrak{S}$ remain simple. For example, the analogue of Eq.~\eqref{eq:fermionfieldmassrelation} at any order $n$ is
\begin{equation}
  \left[(m_i^2-m_j^2)\delta Z^{(n)}_{ji}-m_j\delta m^{(n)}_{ji}-m_i \delta m^{(n)\dagger}_{ji}\right]u_i=-(\cancel{p}+m_j)\breve{\mathfrak{S}}^{(n)}_{ji}u_i\,,
\end{equation}
i.e. the relation keeps exactly the same form at higher orders and we have defined $\mathfrak{S}^{(n)}=\breve{\mathfrak{S}}^{(n)}-\delta m^{(n)}$. Similarly, the Nielsen identity in Eq.~\eqref{eq:Nielsensimple} repeat the form of the 1-loop contributions in Eq.~\eqref{eq:nielsenfirst} at every order. As the form stays the same, the first hint is that the mass structure $m_i^2-m_j^2$ repeats at every order as we have already shown in~\cite{drauksas2023} via a more involved calculation. As the higher orders repeat the structures of the 1-loop computation we expect that $\mathfrak{S}$ and $\mathfrak{L}$ admit the usage of the fine-grained decomposition at least to some extent. If that is the case, it immediately follows that one can promote the 1-loop order definition of the anti-hermitian part of the field renormalization in Eq.~\eqref{eq:ZAfullsimple} to any order by a simple substitution of $\Sigma\to\breve{\mathfrak{S}}$. Simultaneously, one would promote the gauge-independent combinations of Eqs.~\eqref{eq:firstlinedmla}--\eqref{eq:secondlinedmlb} to all orders via the same replacement. However, this remains a hint and beyond the scope of this paper, for example, the fine-grained decomposition for the self-energies can produce terms such as $\Sigma^{X,\times\bullet}$ beyond 1-loop, but at the moment it is not clear whether these cancel out with contributions from lower orders in $\mathfrak{S}$.  

\section{\texorpdfstring{$W$}{W} decay widths with old data}
\label{sec:Bcomparing08}

As mentioned in Section~\ref{sec:numres}, we were unable to fully reproduce the results of~\cite{kniehl2000} and~\cite{almasy2009err}. A discrepancy of similar size with the results of~\cite{kniehl2000} is also reported in~\cite{kara2013}, although, as the authors note, they use different methods to compute the decay width and differences may be expected. In any case, we provide our results for the partial $W$ decay widths at 1-loop with the inputs used by~\cite{almasy2009err} in Tables~\ref{tab:normalschemes08} and~\ref{tab:otherschemes08}, that should serve as an additional reference point and perhaps help with reproducibility. In addition, we provide the numerical value for $\Delta r$ we have found with the inputs of~\cite{almasy2009err}
\begin{equation}
  \Delta r_\text{\cite{almasy2009err}}=0.03835488\,.
\end{equation}

\begin{table}[h!]
\centering
\renewcommand{\arraystretch}{1.3}
\begin{tabular}{|l|c|c|c|c|c|c|}
\hline\hline
\multirow{2}{1.6cm}{\textbf{Partial widths}}
& \multirow{2}{1.9cm}{\centering \textbf{Tree level}}
& \multirow{2}{1.9cm}{\textbf{Ref.~\cite{denner1990}}} 
& \multirow{2}{1.9cm}{\textbf{Ref.~\cite{gambino1999}}} 
& \multirow{2}{1.9cm}{\textbf{Ref.~\cite{dienerkniehl}} }
& \multirow{2}{1.9cm}{\textbf{Ref.~\cite{kniehl2006}} }
& \multirow{2}{1.9cm}{\textbf{Ref.~\cite{kniehl2009a}}}  
\\
&  &  &  &  &  &  \\
\hline

$\bar{\Gamma}^{ud}\times 10$
& 6.473886 & 6.697014 & 6.697014 & 6.697014 &  6.697014 & 6.697014 \\

$\bar{\Gamma}^{us}\times 10^2$
& 3.474831 & 3.594603 & 3.594603 & 3.594603 & 3.594603 & 3.594603\\

$\bar{\Gamma}^{ub}\times 10^6$
& 8.767192 & 9.088798 & 9.088842 & 9.088890 & 9.088842 & 9.088842\\

$\bar{\Gamma}^{cd}\times 10^2$
& 3.469261 & 3.589742 & 3.589742 & 3.589742 & 3.589742 & 3.589742 \\

$\bar{\Gamma}^{cs}\times 10$
& 6.460319 & 6.684691 & 6.684691 & 6.684691 & 6.684691 & 6.684691 \\

$\bar{\Gamma}^{cb}\times 10^3$
& 1.167520 & 1.210657 & 1.210662 & 1.210669 & 1.210662 & 1.210662 \\
\hline
$\bar{\Gamma}^\mathrm{had}$
& 1.364038 & 1.411234 & 1.411234 & 1.411234 & 1.411234 & 1.411234 \\
\hline\hline
\end{tabular}
\caption{Partial and total hadronic $W$ decay widths (in $\mathrm{GeV}$) in various CKM renormalization schemes found in the literature. The numerical inputs are as in~\cite{almasy2009err}.}
\label{tab:normalschemes08}
\end{table}

\begin{table}[h!]
\centering
\renewcommand{\arraystretch}{1.3}
\begin{tabular}{|l|c|c|}
\hline\hline
\multirow{2}{1.6cm}{\textbf{Partial widths}}
& \multirow{2}{2cm}{\centering $\overline{\text{MS}}$ \textbf{scheme}}
& \multirow{2}{2cm}{$ V^\mathrm{ext.}_{ji}=\delta_{ji}$ }
\\
&  &  \\
\hline

$\bar{\Gamma}^{ud}\times 10$
& 6.697010 & 6.697014 \\

$\bar{\Gamma}^{us}\times 10^2$
& 3.594631 & 3.594603 \\

$\bar{\Gamma}^{ub}\times 10^6$
& 9.155930& 9.088842 \\

$\bar{\Gamma}^{cd}\times 10^2$
& 3.589737 & 3.589742 \\

$\bar{\Gamma}^{cs}\times 10$
& 6.684602 & 6.684691 \\

$\bar{\Gamma}^{cb}\times 10^3$
& 1.219597 & 1.210662 \\
\hline
$\bar{\Gamma}^\mathrm{had}$
& 1.411234 & 1.411234 \\
\hline\hline
\end{tabular}
\caption{Partial and total hadronic $W$ decay widths (in $\mathrm{GeV}$) in the $\overline{\mathrm{MS}}$ renormalization scheme for the CKM matrix and a scheme where there is no mixing on external legs (trivial CKM counterterm). The numerical inputs are as in~\cite{almasy2009err}.}
\label{tab:otherschemes08}
\end{table}

\pagebreak{}
\printbibliography

\end{document}